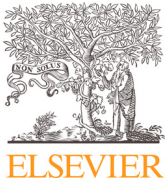
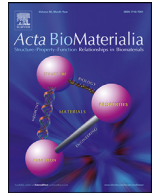

Full length article

# Phase-field modeling of pitting and mechanically-assisted corrosion of Mg alloys for biomedical applications

Sasa Kovacevic [a,*], Wahaaj Ali [b,c], Emilio Martínez-Pañeda [a,*], Javier LLorca [b,d,*]

[a] *Department of Civil and Environmental Engineering, Imperial College London, London SW7 2AZ, UK*
[b] *IMDEA Materials Institute, C/Eric Kandel 2, Getafe 28906, Madrid, Spain*
[c] *Department of Material Science and Engineering, Universidad Carlos III de Madrid, Leganes 28911, Madrid, Spain*
[d] *Department of Materials Science, Polytechnic University of Madrid, E. T. S. de Ingenieros de Caminos, 28040 Madrid, Spain*

## ABSTRACT

A phase-field model is developed to simulate the corrosion of Mg alloys in body fluids. The model incorporates both Mg dissolution and the transport of Mg ions in solution, naturally predicting the transition from activation-controlled to diffusion-controlled bio-corrosion. In addition to uniform corrosion, the presented framework captures pitting corrosion and accounts for the synergistic effect of aggressive environments and mechanical loading in accelerating corrosion kinetics. The model applies to arbitrary 2D and 3D geometries with no special treatment for the evolution of the corrosion front, which is described using a diffuse interface approach. Experiments are conducted to validate the model and a good agreement is attained against *in vitro* measurements on Mg wires. The potential of the model to capture mechano-chemical effects during corrosion is demonstrated in case studies considering Mg wires in tension and bioabsorbable coronary Mg stents subjected to mechanical loading. The proposed methodology can be used to assess the *in vitro* and *in vivo* service life of Mg-based biomedical devices and optimize the design taking into account the effect of mechanical deformation on the corrosion rate. The model has the potential to advocate further development of Mg alloys as a biodegradable implant material for biomedical applications.

**Statement of significance**

A physically-based model is developed to simulate the corrosion of bioabsorbable metals in environments that resemble biological fluids. The model captures pitting corrosion and incorporates the role of mechanical fields in enhancing the corrosion of bioabsorbable metals. Model predictions are validated against dedicated *in vitro* corrosion experiments on Mg wires. The potential of the model to capture mechano-chemical effects is demonstrated in representative examples. The simulations show that the presence of mechanical fields leads to the formation of cracks accelerating the failure of Mg wires, whereas pitting severely compromises the structural integrity of coronary Mg stents. This work extends phase-field modeling to bioengineering and provides a mechanistic tool for assessing the service life of bioabsorbable metallic biomedical devices.

## 1. Introduction

Magnesium (Mg) and its alloys are highly attractive for temporary biomedical implants [1,2]. Good biocompatibility, biodegradability, and mechanical properties place Mg at an advantage over traditional biodegradable polymers for load-bearing applications [3]. Temporary Mg implants are intended to gradually dissolve *in vivo* at a synchronized rate with bone/tissue growth and safely absorb in the human body after healing with no implant residues, thereby avoiding the need for a second operation for implant removal. Those implants have shown promising results in several applications, including orthopedic surgery [4], cardiovascular stents [5], and implants for oral and maxillofacial bones, three-dimensional scaffolds, soft tissue, and nerve regeneration [6].

* Corresponding authors.
*E-mail addresses:* s.kovacevic@imperial.ac.uk (S. Kovacevic), e.martinez-paneda@imperial.ac.uk (E. Martínez-Pañeda), javier.llorca@upm.es, javier.llorca@imdea.org (J. LLorca).

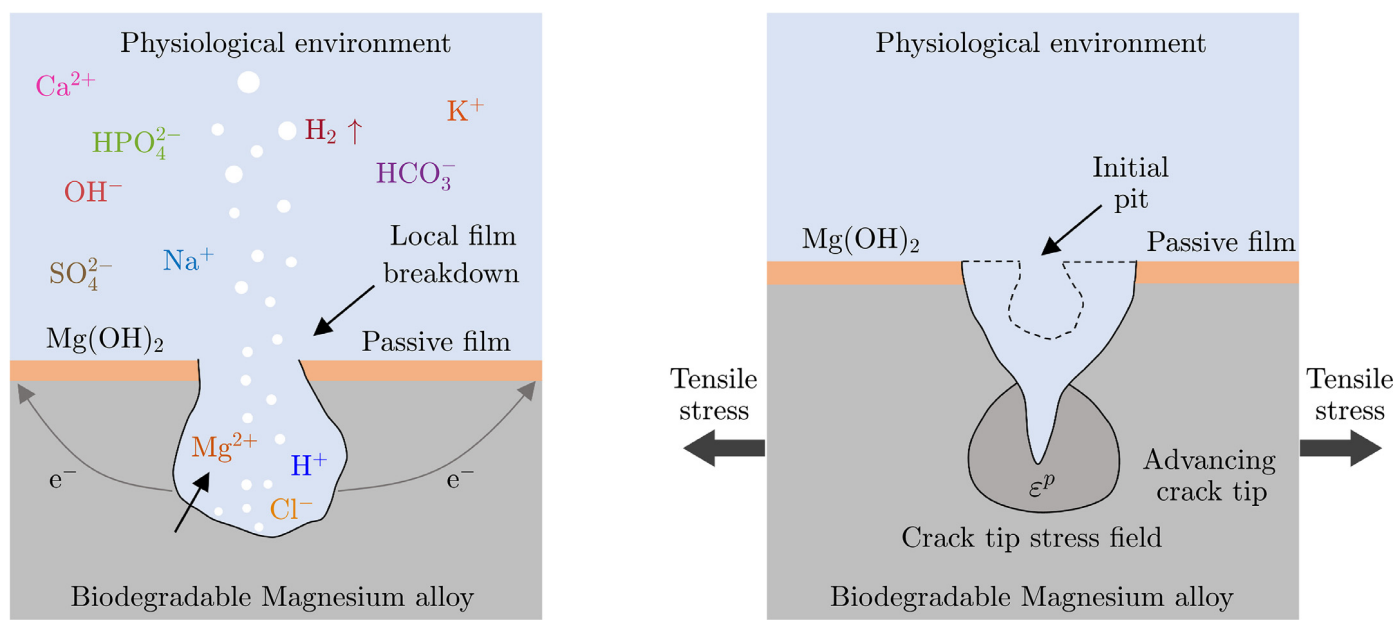

**Fig. 1.** Schematic illustration of the pitting corrosion (left) and stress corrosion cracking (right) mechanisms of Mg alloys during immersion in the physiological environment (simplified representation). Pitting is caused by breakages in the passive film exposing the Mg alloy to the corrosive environment.

Despite successful clinical studies, Mg-based implants have only been approved for a limited number of applications. Rapid corrosion in an aggressive chloride medium like human body fluids is the main reason limiting the widespread use of Mg alloys as a biodegradable material [7]. *In vivo* [8–10] and *in vitro* studies [11–13] have reported that biodegradable Mg alloys (as well as industrially used ones) generally corrode in a localized fashion (e.g., pitting), Fig. 1. For instance, an *in vivo* study [14] has shown that Mg alloy plates and screws in the diaphyseal area of long bones in pigs were nearly completely degraded after 12 months of implantation. The screws showed a faster and nonhomogeneous degradation profile in the intramedullary cavity owing to enhanced exposure to interstitial fluids. Another study [15] has found that corrosion resistance is the major challenge for using Mg interference screws for anterior cruciate ligament reconstruction. Moreover, regions of the interference screw exposed to synovial fluids in the knee joint cavity suffer accelerated degradation.

Different strategies, mainly based on surface modification, have been developed to mitigate these risks and improve the corrosion resistance and biocompatibility of Mg alloys [16,17]. Yet, pitting corrosion could only be diminished to a certain extent [18,19]. Moreover, load-bearing biomedical devices are continuously subjected to various loading conditions during service. In such an environment, biodegradable Mg alloys show sensitivity to stress corrosion cracking (SCC) [19–22]. The synergistic effect of mechanical loading and a corrosive environment significantly reduces the corrosion resistance and mechanical integrity of Mg alloys. The concurrence of these two factors dramatically accelerates the corrosion rate and promotes crack propagation (Fig. 1), leading to the sudden failure of implants [23]. A recent study [24] has reported that elastic strains decrease the corrosion potential, increase corrosion current and accelerate the degradation of WE43 Mg wires, while plastic strains enhance localized corrosion. Hence, SCC is a severe concern for thin implant applications like stents, membranes, and wires.

Before clinical trials, the corrosion performance of Mg alloys is usually determined in *in vitro* tests under various corrosive environments that resemble biological fluids. Ideally, this information can be used to calibrate numerical models to predict the *in vivo* performance of implants and to guide design taking into account their progressive degradation. A wide variety of computational tools have been developed to predict the corrosion behavior of biodegradable Mg alloys. Phenomenological models based on the continuum damage (CD) theory [25] for uniform and localized corrosion [26,27] are based on a scalar damage parameter that reduces the mechanical properties of the corroded regions of the material. The damage evolution law for stress corrosion depends on threshold stress above which corrosion progresses [26] while pitting corrosion is introduced via a dimensionless pitting parameter controlled by a distribution probability density function [27]. The CD approach is further improved by including several different features [28–32]. However, the diffusion process, as the underlying physical mechanism in the corrosion process, is not included. In more advanced phenomenological models [33,34], the diffusion of Mg ions and the evolution of other species have been incorporated through physicochemical interactions and diffusion equations.

Although phenomenological models have been typically used to simulate Mg corrosion, there is a growing interest in the development of physically-based models that can resolve the physical processes governing corrosion and thus provide mechanistic predictions and insight [35–41]. While the underlying physics is relatively well understood, there are significant theoretical and computational challenges intrinsic to the coupled nature of the problem and the difficulty of tracking the evolution of complex corrosion interfaces in arbitrary domains. Regarding the former, a mechanistic model of Mg corrosion must account for Mg dissolution at the corrosion front (short-range interactions), diffusion of Mg ions in solution (long-range interactions) and the coupling with other physical phenomena such as capturing the role of mechanics in enhancing corrosion rates and the electrochemistry-corrosion interplay. Regarding the challenges associated with tracking an evolving corrosion front computationally, a number of numerical techniques have been recently proposed, including Arbitrary Lagrangian-Eulerian (ALE) approaches [35–37], level set methods [38–40], and peridynamics [41]. However, these are mainly used in the context of uniform corrosion as they are still limited in capturing localized corrosion, coupling with other physicomechanical phenomena, and handling geometric interactions in arbitrary di-mensions (2D/3D), such as the coalescence of corrosion pits. Phase-field formulations have emerged as a promising approach for modeling moving interfaces and handling topological changes at different length scales [42]. In phase-field models, the interface between two phases is smoothed over a thin diffuse region using a continuous auxiliary field variable (e.g., $\phi$), see Fig. 2. The phase-field variable $\phi$ has a distinct value in each phase (e.g., 0 and 1), while varying smoothly in between. The movement of the interface is implicitly tracked without presumptions or prescribing the interface velocity. Topological changes of arbitrary complexity (e.g., divisions or merging of interfaces) can be naturally captured in 2D and 3D without requiring any special treatments or ad hoc criteria. The phase-field method has been recently extended to several challenging interfacial phenomena relevant to corrosion in non-biodegradable metallic materials [43–50] and to internal galvanic corrosion and damage localization induced by insoluble secondary phases in Mg alloys [51]. This work aims to extend this success to biomaterial degradation, presenting the first phase-field model for the surface-based localized (pitting) corrosion of biodegradable Mg alloys that incorporates material dissolution, Mg ionic transport, and mechano-chemical interactions.

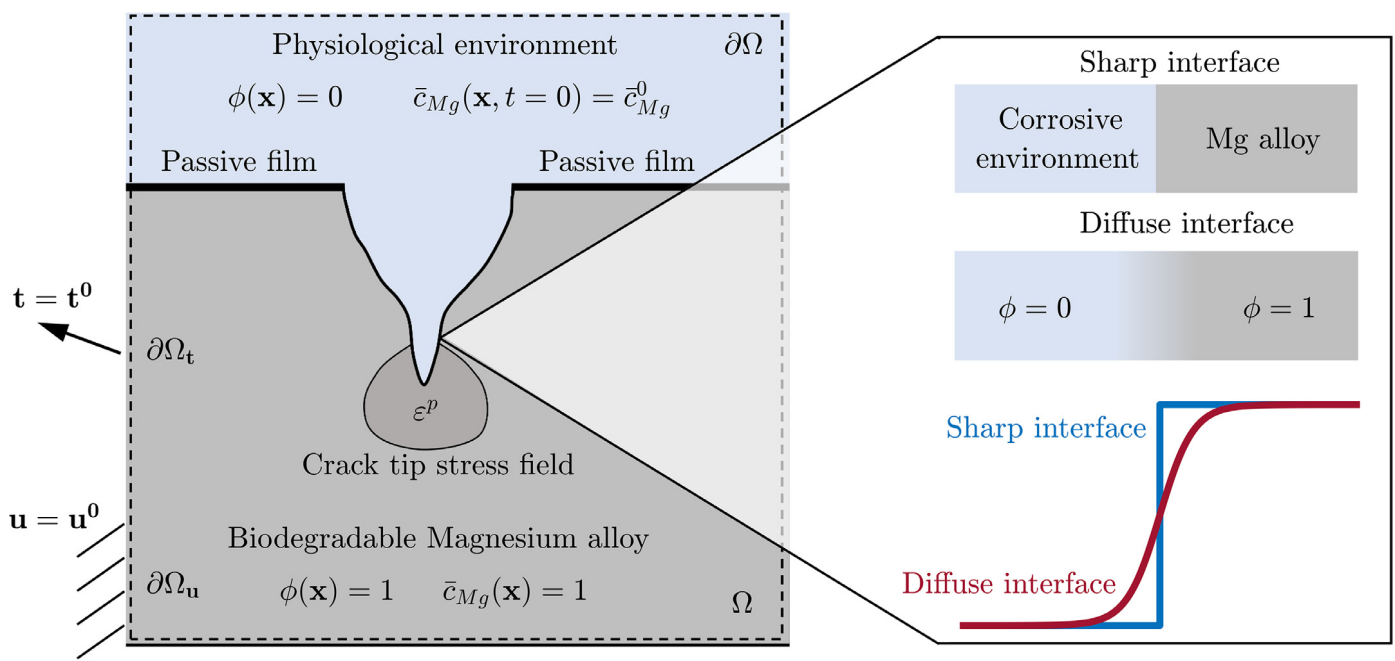

**Fig. 2.** Problem formulation and diffuse interface description of the liquid (physiological environment $\phi = 0$) and solid (biodegradable Mg alloy $\phi = 1$) phases.

The outline of the paper is as follows. The degradation mechanisms governing mechanically-assisted corrosion of biodegradable Mg alloys are presented in the following section and the phase-field model is subsequently formulated. The interplay between Mg dissolution, ionic transport in solution, and mechanical straining is captured by defining a generalized thermodynamic free energy functional that incorporates chemical, interfacial, and mechanical terms. The impact of mechanical fields in accelerating corrosion kinetics is integrated through a mechano-electrochemical mobility term that depends on local stress and strain distributions. The constructed model is calibrated and validated against *in vitro* corrosion data on WE43 Mg alloy wires immersed in simulated body fluid in Section 3. Dedicated experiments are conducted to validate the model, both qualitatively (pitting patterns) and quantitatively (hydrogen gas released), demonstrating as well its ability to capture localized corrosion phenomena. After validation, the potential of the model to handle mechano-chemical effects during corrosion is demonstrated in Section 4 through two representative case studies: pitting corrosion associated with the local failure of a protective layer and the nonhomogeneous stress state of a bioabsorbable coronary stent. The potential of the model is discussed in Section 5 along with recommendations for future work. Conclusions of the investigation are summarized in Section 6.

## 2. The phase-field model for corrosion of Mg

### 2.1. Degradation mechanisms

Magnesium dissolution in aqueous environments, such as biological fluids, is governed by an electrochemical reaction and the corrosion process can be summarized as follows [3]

$$\text{Mg}_{(s)} \rightarrow \text{Mg}^{2+}_{(aq)} + 2e^- \quad \text{(anodic reaction)}$$
$$2\text{H}_2\text{O}_{(aq)} + 2e^- \rightarrow \text{H}_{2(g)} \uparrow + 2\text{OH}^-_{(aq)} \quad \text{(cathodic reaction)}$$
$$\text{Mg}^{2+}_{(aq)} + 2\text{OH}^-_{(aq)} \rightarrow \text{Mg(OH)}_{2(s)} \downarrow \quad \text{(product formation).} \quad (1)$$

The last reaction in Eq. (1) is the precipitation reaction that leads to the formation of a passive layer of magnesium hydroxide (Mg(OH)$_2$) on the Mg surface, Fig. 1. Chloride ions (Cl$^-$) present in the physiological environment react with Mg(OH)$_2$ and transform the protective film into highly soluble magnesium chloride (MgCl$_2$)

$$\text{Mg(OH)}_{2(s)} + 2\text{Cl}^-_{(aq)} \rightarrow \text{MgCl}_{2(aq)} + 2\text{OH}^-_{(aq)} \quad \text{(layer dissolution),} \quad (2)$$

undermining the integrity of the passive film. Fast corrosion rates and pitting corrosion are generally associated with aggressive chloride ions [52]. The presence of inorganic ions and organic compounds in body fluids further increases the complexity of the degradation process [53]. While the effect of certain organic compounds [54] and inorganic ions [55,56] on the corrosion rate have been identified, the degradation mechanisms of Mg alloys in body fluids are not fully understood [57,58]. Therefore, it is assumed that the primary degradation mechanism is driven by the bulk diffusion of Mg ions in the physiological environment. The complex composition of the porous protective layer, its negligible thickness compared to the size of the surrounding body fluid, and the high solubility of MgCl$_2$ in aqueous environments allow to neglect the product formation and layer dissolution reactions, an approach frequently followed in the literature [26,27,32,34,35,41].

The presence of mechanical stresses increases the corrosion susceptibility of Mg alloys. *In vitro* studies [24,59] have indicated that mechanical fields decrease the corrosion potential of Mg alloys, thereby increasing corrosion current densities and dissolution rates. Following Gutman's theory of mechano-electrochemical interactions [60], the anodic dissolution kinetics is given as

$$\frac{i}{i_0} = \left(\frac{\varepsilon^p}{\varepsilon_y} + 1\right) \exp\left(\frac{\sigma_h V_m}{RT}\right), \quad (3)$$

where $i$ is the anodic dissolution current in the presence of mechanical stresses, $i_0$ the anodic dissolution current in the absence of mechanical stresses, $\varepsilon^p$ the effective plastic strain, $\varepsilon_y$ the initial yield strain, $\sigma_h$ the hydrostatic stress, $V_m$ the molar volume of the metal, $R$ the universal gas constant, and $T$ the absolute temperature. After rupture of the passive film and pit nucleation, local stress and plastic strain distributions intensify local material dissolution in the vicinity of the pit, promoting pit-to-crack transition and crack propagation, as schematically illustrated in Fig. 1. As shown below (Section 2.4), the amplification factor in Eq. (3) is embedded into the model kinetics parameter that characterizes solid-liquid interface movement to incorporate the role of mechanical fields in accelerating corrosion.

### 2.2. Thermodynamics

The problem formulation is depicted in Fig. 2 and could be summarized as follows. The system consists of a biodegradable Mg alloy in contact with physiological environments that, by composition, mimic body fluids. The system domain $\Omega$ includes both the Mg alloy and the corrosive environment. A continuous phase-field parameter $\phi$ is introduced to distinguish different phases: $\phi = 1$ represents the solid phase (Mg alloy), $\phi = 0$ corresponds to the liquid phase (physiological fluid), and $0 < \phi < 1$ indicates the thin interfacial region between the phases (solid-liquid interface). With vanishing normal fluxes ($\mathbf{n} \cdot \mathbf{J} = 0$) on the domain boundary $\partial \Omega$, the independent kinematic variables necessary for model description are the non-conserved phase-field parameter describing the evolution of the corroding interface $\phi(\mathbf{x}, t)$, the displacement vector to characterize deformation of the solid phase $\mathbf{u}(\mathbf{x}, t)$, and the normalized concentration of Mg ions $\bar{c}_{Mg}(\mathbf{x}, t)$ with respect to the concentration in the solid phase ($\bar{c}_{Mg} = c_{Mg}/c^s_{Mg}$). More details regarding nondimensionalization are given in Section 2.5.

The free energy functional for a heterogeneous system such as the one in Fig. 2 can be written as

$$\mathcal{F} = \int_\Omega \left[ f^{chem}(\bar{c}_{Mg}, \phi) + f^{grad}(\nabla \phi) + f^{mech}(\nabla \mathbf{u}, \phi) \right] d\Omega, \quad (4)$$

where $f^{chem}$, $f^{grad}$, and $f^{mech}$ are the chemical, gradient, and mechanical energy densities defined below.

#### 2.2.1. Chemical free energy density

Following the phase-field model for phase transitions in binary alloys [61], the chemical free energy density of a homogeneous system consisting of solid and liquid phases is decomposed into the chemical energy density associated with material composition and double-well potential energy

$$f^{chem}(\bar{c}_{Mg}, \phi) = (1 - h(\phi)) f_l^{chem}(\bar{c}^l_{Mg}) + h(\phi) f_s^{chem}(\bar{c}^s_{Mg}) + \omega g(\phi), \quad (5)$$

where $f_l^{chem}(\bar{c}^l_{Mg})$ and $f_s^{chem}(\bar{c}^s_{Mg})$ are the chemical free energy densities within the liquid and solid phases as a function of normalized phase-concentrations $\bar{c}^l_{Mg}$ and $\bar{c}^s_{Mg}$. In the above equation, $g(\phi)$ and $h(\phi)$ are the double-well potential energy and interpolation functions commonly expressed as

$$\begin{aligned} g(\phi) &= 16\phi^2(1-\phi)^2 \\ h(\phi) &= \phi^3\left(6\phi^2 - 15\phi + 10\right). \end{aligned} \quad (6)$$

$\omega$ in Eq. (5) is a constant that determines the energy barrier height at $\phi = 1/2$ between the two minima at $\phi = 0$ and $\phi = 1$.

The chemical free energy densities within each phase in Eq. (5) are approximated by simple parabolic functions with the same curvature parameter $A$ as

$$\begin{aligned} f_l^{chem}\left(\bar{c}^l_{Mg}\right) &= \frac{1}{2} A \left(\bar{c}^l_{Mg} - \bar{c}^{l,eq}_{Mg}\right)^2 \\ f_s^{chem}\left(\bar{c}^s_{Mg}\right) &= \frac{1}{2} A \left(\bar{c}^s_{Mg} - \bar{c}^{s,eq}_{Mg}\right)^2, \end{aligned} \quad (7)$$

where $\bar{c}^{l,eq}_{Mg} = c^{l,eq}_{Mg}/c^s_{Mg}$ and $\bar{c}^{s,eq}_{Mg} = c^{s,eq}_{Mg}/c^s_{Mg}$ are the normalized equilibrium Mg concentrations in the liquid and solid phases (refer to Section 2.5 for dimensional analysis). Alternatively, the chemical free energy density can be approximated assuming a dilute solution [46–48]. Physically, the equilibrium concentration in the solid phase $c^{s,eq}_{Mg}$ represents the average concentration of Mg ions within the material. Since the product formation and protective layer dissolution are neglected in the current work (Section 2.1), $c^{l,eq}_{Mg}$ is determined based on the mass density and molar mass of MgCl$_2$ formed on the exposed Mg surface.

The interfacial region is defined as a mixture of both phases with different concentrations but with the same diffusion chemical potential [61]

$$\begin{aligned} \bar{c}_{Mg} &= (1 - h(\phi))\bar{c}^l_{Mg} + h(\phi)\bar{c}^s_{Mg} \\ \frac{\partial f_l^{chem}\left(\bar{c}^l_{Mg}\right)}{\partial \bar{c}^l_{Mg}} &= \frac{\partial f_s^{chem}\left(\bar{c}^s_{Mg}\right)}{\partial \bar{c}^s_{Mg}}. \end{aligned} \quad (8)$$

Using Eqs. (7) and (8) renders the following definition for the chemical free energy density of the system

$$f^{chem}(\bar{c}_{Mg}, \phi) = \frac{1}{2} A \left[\bar{c}_{Mg} - h(\phi)(\bar{c}^{s,eq}_{Mg} - \bar{c}^{l,eq}_{Mg}) - \bar{c}^{l,eq}_{Mg}\right]^2 + \omega g(\phi). \quad (9)$$

#### 2.2.2. Gradient energy density

The interfacial energy density is defined as

$$f^{grad}(\nabla \phi) = \frac{1}{2}\kappa |\nabla \phi|^2, \quad (10)$$

where $\kappa$ is the isotropic gradient energy coefficient. The phase-field parameters $\omega$ and $\kappa$ are connected to the physical quantity (interfacial energy $\Gamma$) and computational parameter (interface thickness $\ell$). For the accepted double well potential $g(\phi)$ in Eq. (6), the following relations are obtained [62]

$$\omega = \frac{3\Gamma}{4\ell} \qquad \kappa = \frac{3}{2}\Gamma\ell. \quad (11)$$

#### 2.2.3. Strain energy density

The mechanical behavior of the solid phase is assumed to follow the von Mises theory of plasticity [63]. Considering deformable elasto-plastic solids, the mechanical free energy density $f^{mech}$ in Eq. (4) is additively decomposed into elastic $f_e^{mech}$ and plastic components $f_p^{mech}$

$$f^{mech}(\nabla \mathbf{u}, \phi) = h(\phi)(f_e^{mech} + f_p^{mech}), \quad (12)$$

where $h(\phi)$ ensures the transition from the intact solid (uncorroded Mg alloy) to the completely corroded (liquid) phase. The elastic strain energy density $f_e^{mech}$ is a quadratic form of the elastic strain

$$f_e^{mech}(\nabla \mathbf{u}) = \frac{1}{2}\boldsymbol{\varepsilon}^e : \mathbf{C} : \boldsymbol{\varepsilon}^e \qquad \boldsymbol{\varepsilon}^e = \boldsymbol{\varepsilon} - \boldsymbol{\varepsilon}^p, \tag{13}$$

where $\mathbf{C}$ is the rank-four elastic stiffness tensor and $\boldsymbol{\varepsilon}^e$ is the elastic strain tensor obtained by subtracting the plastic strain tensor $\boldsymbol{\varepsilon}^p$ from the total strain $\boldsymbol{\varepsilon}$. For linearized kinematics, the total strain tensor is the symmetric part of the displacement gradient

$$\boldsymbol{\varepsilon} = \frac{1}{2}(\nabla \mathbf{u} + (\nabla \mathbf{u})^T). \tag{14}$$

The elastic deformation of the solid is described by the isotropic linear elasticity theory so that the rank-four elastic stiffness tensor reads

$$C_{ijkl} = \lambda \delta_{ij}\delta_{kl} + \mu(\delta_{ik}\delta_{jl} + \delta_{il}\delta_{jk}), \tag{15}$$

where $\lambda$ and $\mu$ are the Lamé elastic constants.

The plastic strain energy density $f_p^{mech}$ is incrementally computed from the plastic strain tensor $\boldsymbol{\varepsilon}^p$ and the Cauchy stress tensor $\boldsymbol{\sigma_0}$ for the intact configuration as

$$f_p^{mech} = \int_0^t \boldsymbol{\sigma_0} : \dot{\boldsymbol{\varepsilon}}^p \, dt. \tag{16}$$

### 2.3. Governing equations

Using the balance of power and the principle of virtual power [64], the following time-dependent governing equations for the independent kinematic fields $\phi(\mathbf{x},t)$, $\bar{c}_{Mg}(\mathbf{x},t)$, and $\mathbf{u}(\mathbf{x},t)$ are derived (see details in the Supplementary Materials)

$$\begin{cases} \frac{\partial \phi}{\partial t} = -L\left(\frac{\partial f^{chem}}{\partial \phi} - \kappa \nabla^2 \phi\right) \\ \frac{\partial \bar{c}_{Mg}}{\partial t} = -\nabla \cdot \mathbf{J} \\ \mathbf{J} = -D_{c_{Mg}}\nabla \bar{c}_{Mg} - D_{c_{Mg}}h'(\phi)\left(\bar{c}_{Mg}^{l,eq} - \bar{c}_{Mg}^{s,eq}\right)\nabla \phi \\ \nabla \cdot \boldsymbol{\sigma} = \mathbf{0} \end{cases} \text{in } \Omega, \tag{17}$$

complemented with boundary conditions

$$\begin{cases} \kappa \mathbf{n} \cdot \nabla \phi = 0 \text{ and } \mathbf{n}\cdot\mathbf{J}=0 \text{ on } \partial \Omega \\ \mathbf{t} = \mathbf{n}\cdot\boldsymbol{\sigma} = \mathbf{t}^0 \text{ on } \partial\Omega \text{ and } \mathbf{u} = \mathbf{u}^0 \text{ on } \partial \Omega_u \end{cases}. \tag{18}$$

The resulting set of governing equations includes the Allen-Cahn equation [65] for the non-conserved phase-field parameter, the diffusion equation for the Mg concentration in the liquid and solid phases, and the linear momentum balance equation for quasi-static mechanical deformation. In the above equation, $L$ is the kinetic coefficient that characterizes the interfacial mobility and $D_{c_{Mg}}$ the effective diffusion coefficient interpolated with the phase-field parameter between the phases

$$D_{c_{Mg}} = D^s_{c_{Mg}}h(\phi) + (1 - h(\phi))D^l_{c_{Mg}}, \tag{19}$$

where $D^l_{c_{Mg}}$ and $D^s_{c_{Mg}}$ stand for the diffusion coefficients of Mg ions in the liquid (corrosive environment) and solid phases. $D^s_{c_{Mg}} \ll D^l_{c_{Mg}}$ is enforced to retard diffusion of Mg ions inside the solid phase. The role of mechanical fields on the interface kinetics is incorporated by modifying the interface mobility parameter $L$, which includes a mechano-electrochemical contribution that amplifies the dissolution process, as shown in Section 2.4. Thus, the mechanical term $\partial f^{mech}/\partial \phi = h'(\phi)f^{mech}$ is neglected in the phase-field equation (Eq. (17)). For an alternative way of incorporating the me-chanical contribution to interface kinetics, the interested reader is referred to Refs. [46–48].

### 2.4. Mechano-electrochemical coupling

The role of mechanical fields in enhancing corrosion kinetics is incorporated by following Gutman's theory [60]. As shown in Eq. (3), the anodic dissolution can be amplified by an amplification factor that depends on local stress and strain distributions. As the anodic dissolution kinetics dictates interface motion, the interfacial mobility coefficient $L$ is analogously connected to mechanical fields. Using Eq. (3) and considering the linear relationship between $L$ and $i_a$ (corrosion current density) [45] returns the following expression for the kinetic coefficient in Eq. (17)

$$\frac{L}{L_0} = \left(\frac{\varepsilon^p}{\varepsilon_y} + 1\right)\exp\left(\frac{\sigma_h V_m}{RT}\right), \tag{20}$$

where $L_0$ is the interfacial mobility that physically corresponds to the anodic dissolution current $i_0$ in the absence of mechanical stresses and plastic strains, Eq. (3). The interfacial mobility $L_0$ is determined in Section 3 considering stress-free corrosion experiments on Mg wires.

### 2.5. Dimensional analysis

To facilitate numerical simulations and improve convergence, the governing equations Eq. (17) are normalized using the interface thickness $\ell$ as the characteristic length, Mg concentration in the solid phase $c^s_{Mg}$, diffusion coefficients of Mg ions in the liquid phase $D^l_{c_{Mg}}$, and the energy barrier height $\omega$ as the energy normalization factor. Thus, the nondimensional time $\bar{t}$, nondimensional space coordinates $\bar{\mathbf{x}}$, and nondimensional gradient $\bar{\nabla}$ are given as

$$\bar{t} = \frac{tD^l_{c_{Mg}}}{\ell^2} \qquad \bar{\mathbf{x}} = \frac{\mathbf{x}}{\ell} \qquad \bar{\nabla} = \ell \nabla. \tag{21}$$

Other dimensionless fields and parameters are

$$\begin{aligned} \bar{c}_{Mg} &= c_{Mg}/c^s_{Mg} & \bar{c}^{l,eq}_{Mg} &= c^{l,eq}_{Mg}/c^s_{Mg} \\ \bar{c}^{s,eq}_{Mg} &= c^{s,eq}_{Mg}/c^s_{Mg} & \bar{D}_{c_{Mg}} &= D_{c_{Mg}}/D^l_{c_{Mg}} \\ \bar{f}^{chem} &= f^{chem}/\omega & \bar{\boldsymbol{\sigma}} = \boldsymbol{\sigma}/\omega & \bar{\kappa} = \kappa/(\omega\ell^2). \end{aligned} \tag{22}$$

The above nondimensional variables return the following governing equations

$$\begin{cases} \frac{\partial \phi}{\partial \bar{t}} = -\tau\left(\frac{\partial \bar{f}^{chem}}{\partial \phi} - \bar{\kappa}\bar{\nabla}^2 \phi\right) \\ \frac{\partial \bar{c}_{Mg}}{\partial \bar{t}} = \bar{\nabla} \cdot \left[\bar{D}_{c_{Mg}}\bar{\nabla}\bar{c}_{Mg} + \bar{D}_{c_{Mg}}h'(\phi)\left(\bar{c}^{l,eq}_{Mg} - \bar{c}^{s,eq}_{Mg}\right)\bar{\nabla}\phi\right] \\ \bar{\nabla}\cdot \bar{\boldsymbol{\sigma}} = \mathbf{0} \end{cases} \text{in } \Omega, \tag{23}$$

along with the corresponding nondimensional boundary conditions. The details of the numerical implementation of Eq. (23) are given in Supplementary Materials.

The characteristic times for diffusion $t_d$ and interface reaction $t_\phi$ are then given by

$$t_d = \frac{\ell^2}{D^l_{c_{Mg}}} \qquad t_\phi = \frac{1}{L\omega}, \tag{24}$$

and their ratio

$$\tau = \frac{t_d}{t_\phi} = \frac{L}{D^l_{c_{Mg}}}\ell^2\omega, \tag{25}$$

determines the rate-limiting process. For the case of $\tau \gg 1$ (i.e., $t_d \gg t_\phi$), diffusion is slower than interface reactions and the process is driven by bulk diffusion. This situation is denominated diffusion-controlled corrosion. On the contrary, diffusion is faster if $\tau \ll 1$ (i.e., $t_d \ll t_\phi$) so that there is no accumulation of Mg ions at the metal-fluid interface. Under that condition, the rate of material transport is interface reaction-controlled (commonly called activation-controlled corrosion). The criterion for the interfacial mobility coefficient for the two rate-limiting processes reads

$$L \gg \frac{D^l_{c_{Mg}}}{\ell^2 \omega} \text{ (diffusion-controlled)}$$

$$L \ll \frac{D^l_{c_{Mg}}}{\ell^2 \omega} \text{ (activation-controlled)}. \quad (26)$$

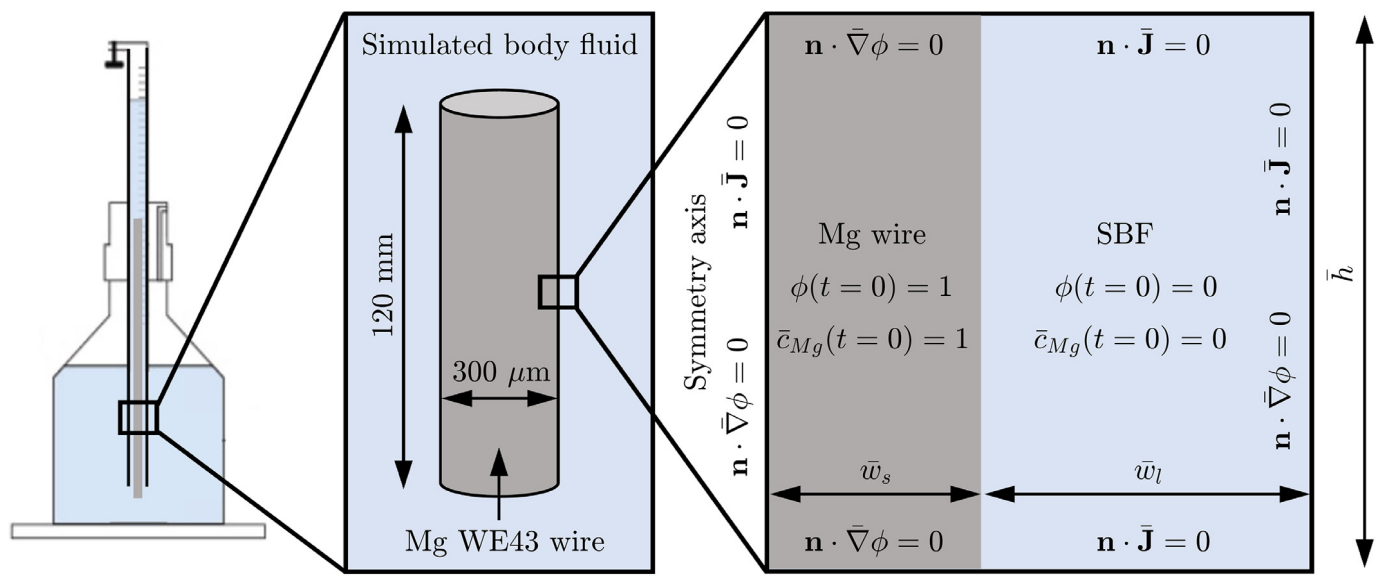

**Fig. 3.** Schematic disposition of the experimental setup (left) and the corresponding nondimensional computational domain (right) for WE43 Mg alloy wires immersed in SBF. The size of the nondimensional computational domain ($\bar{w}_s = 37.5$, $\bar{w}_l = 362.50$, and $\bar{h} = 250$) is normalized using the interface thickness $\ell = 4$ μm as the characteristic length.

The effects of diffusion- and activation-controlled processes on the corrosion behavior are discussed in Section 5.

## 3. Experiments and model validation

### 3.1. Experimental section

#### 3.1.1. Materials and experimental methods

An *in vitro* degradation study was carried out on WE43MEO Mg alloy wires of 0.3 mm in diameter to validate the proposed phase-field model. The Mg wires were manufactured by cold drawing at Meotec GmbH (Aachen, Germany) from WE43MEO Mg alloy with a nominal composition of 1.4–4.2% Y, 2.5–3.5% Nd, <1% (Al, Fe, Cu, Ni, Mn, Zn, Zr) and balance Mg (in wt. %). The wires were annealed at 450 °C for 5 s after cold drawing to reduce the dislocation density induced during drawing and improve the ductility [66,67].

Corrosion tests were carried out in wires of 120 mm in length immersed in Simulated Body Fluid (c-SBF) at 37 °C. The experimental setup is schematically depicted in Fig. 3. The composition (per liter) of the c-SBF was 8.035 g NaCl, 0.355 g NaHCO$_3$, 0.225 g KCl, 0.176 g K$_2$HPO$_4$, 0.145 g MgCl$_2$, 0.292 g CaCl$_2$, 0.072 g Na$_2$SO$_4$, and 50 mL of Tris buffer pH 7.5. The ratio of c-SBF volume to the wire surface area was > 0.5 mL/mm$^2$, according to the ASTM G31-72 standard.

The degradation rate was assessed by measuring the amount of hydrogen gas released which is, according to Eq. (1), equivalent to the mass loss of corroded Mg. To measure the evolved hydrogen gas, the Mg wires were placed inside a glass burette, as illustrated in Fig. 3. The burette was inserted into a sealed plastic bottle filled with SBF. The released hydrogen gas was captured in the burette and tracked by a eudiometer. Twelve samples were used in the immersion tests. After 24 h of immersion, four samples were taken out of the SBF to assess the extent of pitting corrosion. To this end, twenty random cross-sections of the corroded Mg wires were mounted in an epoxy resin, grounded, polished, and images were taken in an optical microscope. These images were analyzed using the PitScan Framework [68] to quantify the degree of pitting corrosion.

#### 3.1.2. Statistical analysis

The experimental measurements and numerically obtained results in Section 3.2 are expressed as mean value ± standard deviation. Microsoft Excel software was used for the statistical calculations.

### 3.2. Model validation

Two types of simulations are performed to validate the phase-field model. First, experimental measurements of hydrogen release over the immersion time are used to calibrate the model kinematic parameter $L_0$ considering uniform corrosion. Second, the wire cross-sections measured after 24 h of immersion in SBF are compared with pitting corrosion predictions. In these simulations, the mechanical effect is not considered. The properties of the Mg alloy used in the simulations are listed in Table 1. Due to the lack of experimental data on the diffusivity of Mg ions in biological fluids, the magnitude of $D^l_{c_{Mg}}$ is estimated as the average value utilized in various numerical studies [33–40]. Although the concentration of Mg ions in the solid phase ($c^s_{Mg}$) could be evaluated using the material data for pure Mg and the mass fraction of alloying elements and impurities, the value for pure Mg is used in this investigation for simplicity [69]. The physiological environment and the presence of chloride ions determine the equilibrium concentration of Mg ions in the liquid phase $c^{l,eq}_{Mg}$ (saturated concentration). As the formation of the partly protective layer is neglected (Section 2.1), the saturated concentration of Mg ions in the cor-rosive environment is calculated based on the mass density and molar mass of MgCl$_2$ formed on the exposed Mg surface. Assuming that the mass density of MgCl$_2$ is 54.20 g/L and its molar mass of 95.21 g/mol, the equilibrium concentration of Mg ions is $c^{l,eq}_{Mg} = 0.57$ mol/L. The role of the saturated concentration in the degradation process is further addressed in Section 5.

**Table 1**
Parameters common to all phase-field simulations.

| Quantity | Value | Unit |
| --- | --- | --- |
| Diffusion coefficient of Mg ions in the liquid phase $D^l_{c_{Mg}}$ | $10^{-10}$ | m$^2$/s |
| Diffusion coefficient of Mg ions in the solid phase $D^s_{c_{Mg}}$ | $10^{-13}$ | m$^2$/s |
| Equilibrium concentration in the liquid phase $c^{l,eq}_{Mg}$ | 0.57 | mol/L |
| Equilibrium concentration in the solid phase $c^{s,eq}_{Mg}$ | 71.44 | mol/L [69] |
| Molar volume of Mg $V_m$ | 13.998 | cm$^3$/mol [69] |
| Interfacial energy $\Gamma$ | 0.5 | J/m$^2$ |
| Interface thickness $\ell$ | 4 | μm |
| Chemical free energy density curvature parameter $A$ | $6.10^7$ | J/m$^3$ |
| Absolute temperature $T$ | 310.15 | K |

The phase-field parameters, energy gradient coefficient $\kappa$ and energy barrier height $\omega$, are connected to the interfacial energy $\Gamma$ and the interface thickness $\ell$, Eq. (11). While the interface thickness is a purely computational parameter whose choice is based on the scale of the problem, the interfacial energy is a physical quantity that depends on crystallographic orientation. The average value reported for pure Mg in Refs. [70,71] is used in this investigation. Lastly, the chemical free energy density curvature parameter $A$ in Eq. (7) is assumed to have a similar value as in Refs. [43,45] for corrosion in metallic materials.

*3.2.1. Phase-field simulations of uniform corrosion*

The phase-field simulations of uniform corrosion are performed using an axisymmetric domain as illustrated in Fig. 3. The nondimensional form of governing Eq. (23) is solved with accompanying initial and boundary conditions. A smooth equilibrium phase-field profile is prescribed as the initial solid-liquid interface. The interface thickness $\ell$ is selected to be significantly smaller than the diameter of the Mg wire ($\ell = 4$ μm). No flux boundary conditions for diffusion and phase-field are imposed at all the outer edges of the domain to simulate an unbounded environment. These boundary conditions preserve mass conservation and imply that no diffusion occurs across the domain boundary. The applied boundary conditions and the large domain size of SBF in the horizontal direction are selected to mimic the experimental setup and ensure that the solution does not saturate with Mg ions. The solid material and the surroundings are isotropic. Thus, the vertical dimension of the domain does not influence the results, and the analysis can be reduced to a one-dimensional axisymmetric problem.

The volume of hydrogen released per unit of exposed area is calculated in the simulations from the ideal gas law

$$H_{gas} = \frac{\Delta n_{Mg} RT}{PA}, \qquad (27)$$

where $P$ is the pressure (1 atm), $A$ the exposed area, and $\Delta n_{Mg}$ the total amount of dissolved Mg in (mol) determined as

$$\Delta n_{Mg} = \int_{\Omega(t)} c_{Mg} d\Omega - \int_{\Omega(t=0)} c_{Mg} d\Omega. \qquad (28)$$

The predicted hydrogen gas evolution per unit area of the Mg wire is plotted as a function of the immersion time in Fig. 4, together with the experimental data obtained from the corrosion tests in c-SBF. The experimental results show that the corrosion rate was initially fast and approximately linear up to 24 h. The corrosion rate slowed down afterward to reach a plateau at 120 h. The phase-field simulations return the same trends and accurately reproduce the experimental data for the hydrogen release using an interfacial mobility parameter $L_0 = 2.3 \cdot 10^{-10}$ m$^3$/(J·s) that corresponds to the $\tau$ value of $3.45 \cdot 10^{-6}$, indicating an activation-controlled process. The decrease in corrosion rate was attributed experimentally to the reduction in surface area with the progress of corrosion due to the circular shape of the wire and the formation of a protective layer, mainly formed by magnesium hydroxide with precipitates of carbonates and phosphates, which hindered the diffusion of SBF solution toward the core of the uncorroded Mg wire [15,67,72,73]. The good agreement between experiments and simulations indicates that the first factor is dominant while the effect of the protective layer (that is not considered in the phase-field simulations) can be neglected in the presence of Cl$^-$ ions.

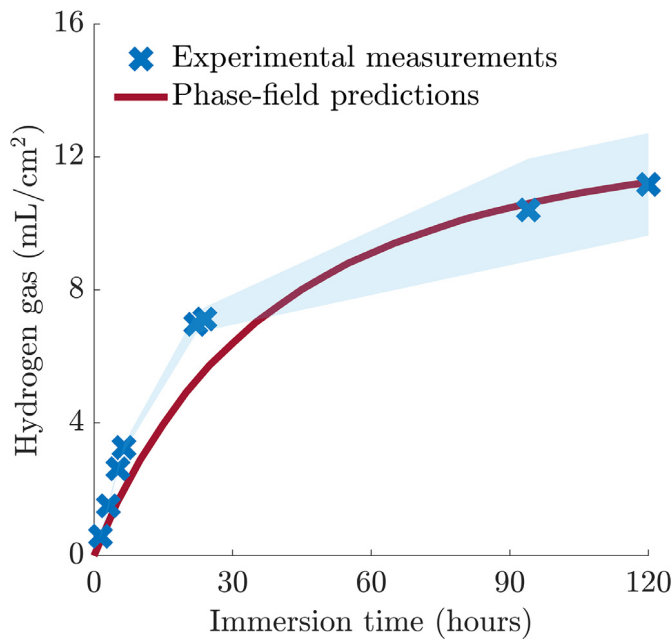

**Fig. 4.** Hydrogen gas evolution as a function of immersion time for Mg wires. Numerical results assuming uniform corrosion and experimental measurements. The light blue area stands for the standard deviation of the experiments.

*3.2.2. Phase-field simulations of pitting corrosion*

A certain degree of randomness needs to be introduced in the system to simulate pitting corrosion. Even assuming uniform alloy composition and surface properties, pitting may occur due to nonuniform distributions of aggressive Cl$^-$ ions, as those ions undermine the protective layer (Eq. (2)). Following that analogy, pitting is introduced in the model through a spatially-dependent kinetic coefficient $L'_0$, correlating it to a random nonuniform distribution of Cl$^-$ ions. Areas with higher values of $L'_0$ reflect the higher concentration of aggressive Cl$^-$ ions, thereby promoting pit-ting corrosion. Random distribution functions are used to define the nonuniform distribution of Cl⁻ ions and to capture the stochastic nature of pitting.

Introducing randomness in terms of the nonuniform distribution of Cl⁻ ions breaks the axial symmetry conditions and, consequently, requires 3D simulations. Performing such simulations for the geometry considered (very long and thin wires) is computationally expensive. Hence, without loss of generality, 3D simulations are replaced by multiple 2D simulations to provide statistical information about the pitting corrosion metrics that can be compared to the experimental data obtained from numerous cross-sections examined in the optical microscope. Space-dependent 2D data is constructed using a sum of trigonometric functions combined with two uniform random distribution functions that introduce randomness in the system. The sum of trigonometric functions can be seen as the assembly of many spatial waves whose amplitudes and phase angles are defined through the two random distribution functions. Therefore, the spatially dependent interfacial mobility parameter $L_0'$ can be written as

$$L_0' = L_0 f(\bar{x}, \bar{y}) = L_0 \left[ a_0 \sum_{m=-N}^{N} \sum_{n=-N}^{N} \frac{\gamma(m,n)}{(m^2+n^2)^{\beta/2}} \cos\left(2\pi(m\bar{x}+n\bar{y}) + \varphi(m,n)\right) + a_1 \right], \quad (29)$$

where $L_0$ is the interfacial mobility parameter determined for uniform corrosion and $f(\bar{x}, \bar{y})$ acts as a dimensionless pitting function. This function is represented by the double summation term (i.e., the sum of spacial waves in the $x$ and $y$ directions) and serves as a stochastic function to introduce randomness in the system. In the previous expression, $\bar{x}$ and $\bar{y}$ are normalized spatial coordinates, $m$ and $n$ are spatial waves in the $x$ and $y$ directions. The number of spatial waves in both directions is $N$. The first uniform random distribution function $\gamma(m,n)$ is defined between zero and one and determines random amplitudes. High-frequency amplitudes are attenuated with the exponent $\beta$ to generate smooth amplitude coefficients. Higher $\beta$ values return a smooth (more uniform) pitting function. The second uniform random distribution function, which controls the phase angle of each wave $\varphi(m,n)$, is defined between $-\alpha\pi/2$ and $\alpha\pi/2$ and governs the spatial distribution of the data. Thus, different spatial distributions, periodicity, magnitudes, and smoothness of random data are controlled with the number of spatial frequencies $N$, exponent $\beta$, and the range of the distribution function $\varphi(m,n)$ varying the $\alpha$ value. For the purpose of pitting corrosion, the coefficients $a_0$ and $a_1$ are included to preserve the non-negativity of the interfacial mobility parameter ($L_0' > 0$) and control the desired difference between the maximum and minimum amplitude to manage the pitting intensity.

Three pair sets of $N$ and $\beta$ are selected, such that the first pair is $N = 2$ and $\beta = 0.1$, the second pair $N = 2.5$ and $\beta = 1.5$, and the third pair $N = 1.25$ and $\beta = 0.75$. For each $N - \beta$ pair set, three different values of the $\alpha$ parameter are considered, i.e., $\alpha = 1$, $\alpha = 3$, and $\alpha = 5$. For each of these nine combinations, three different amplitudes are applied (by adjusting the coefficients $a_0$ and $a_1$) such that the spatially dependent interfacial mobility parameter lies in between $0.5L_0 \leq L_0' \leq 2L_0$, $0.2L_0 \leq L_0' \leq 5L_0$, and $0.1L_0 \leq L_0' \leq 10L_0$. All the other model parameters are identical to those used in the uniform corrosion case. Hence, twenty-seven 2D pitting simulations are carried out to analyze pitting corrosion.

The spatial distribution of the mobility parameter $L_0'$ for $N = 2$, $\beta = 0.1$, and three different $\alpha$ values (1, 3, and 5) is depicted in Fig. 5(a)–(c). The corresponding 2D contour plots of the remaining cross-section of Mg after 24 h of immersion in SBF are given in Fig. 5(d)–(f). Pitting corrosion initiates and follows regions with high $L_0'$ values (i.e., more Cl⁻ ions). Three representative experimental cross-sections of the Mg wires after 24 h of immersion in SBF are plotted in Fig. 5(g)–(i) for the sake of comparison. They are very similar to the phase-field simulations but quantitative comparisons between experiments and simulations can be carried out through three different metrics parameters [68]. They are (i) the uniform corrosion radius in Fig. 6(a) (the radius of the circular section that has the same area as the corroded cross-section), (ii) the average pit depth in Fig. 6(b) (average distance from the degraded cross-section to the uniform corrosion circle), and (iii) the maximum pit depth in Fig. 6(c) (maximum distance from the corroded cross-section to the uniform corrosion circle). The experimental uniform corrosion radius after 24 h of immersion in SBF was 108 ± 21 μm, which corresponds to approximately 48% mass loss and is in agreement with hydrogen gas evolution tests. The higher standard deviation indicates the variation of mass loss among different sections of wire. The experimental values of the average and maximum pit depth were 26 ± 12 μm and 56.62 ± 19.3 μm, which shows the severity of pitting corrosion at particular cross sections. The average experimental values with standard deviations of these three parameters obtained from the analysis of ten different cross-sections are shown in Fig. 6(a)–(c), together with corresponding results obtained from the twenty-seven phase-field simulations. The agreement between experiments and simulations in terms of the uniform corrosion radius and the average pit depth is satisfactory, while the simulations slightly underestimate the maximum pit depth. Overall, the agreement between the experimental measurements and phase-field predictions for hydrogen gas evolution and pitting metrics indicates that the proposed model can be utilized to simulate uniform and pitting corrosion of biodegradable Mg alloys immersed in physiological environments. The model satisfactorily predicts hydrogen gas evolution and captures the experimental trend for pitting corrosion. In the following section, the model is used to ascertain the role of mechanical fields in accelerating the corrosion process.

## 4. Applications

The proposed framework to predict the degradation of Mg alloys in physiological environments is applied in this section to assess the evolution of corrosion in the presence of mechanical stresses in two different scenarios. The first deals with a wire loaded in tension in which the protective layer is damaged, leading to the formation of a pit. The second one analyzes the corrosion of a bioabsorbable Mg alloy coronary stent.

### 4.1. Stress-assisted corrosion of Mg wires

In this simulation, the Mg wire is simultaneously immersed in SBF and subjected to tensile deformation along the wire axis. It is further assumed that the wire surface is protected against corrosion by a thin surface layer locally damaged in a small area. The initial breakdown of the protective layer enables the ingress of aggressive Cl⁻ ions leading to the nucleation of a pit that acts as a stress concentrator. The initial pit has a semi-circular shape with a radius of 10 μm around the whole diameter of the wire to maintain axisymmetric boundary conditions, Fig. 7.

Due to symmetry, only half of the axisymmetric domain is considered in the simulation, as depicted in Fig. 7. Similarly to the previous case, to represent an unbounded domain, no flux (Neumann) boundary conditions are enforced at all the outer boundaries of the computational domain for both the phase-field and the

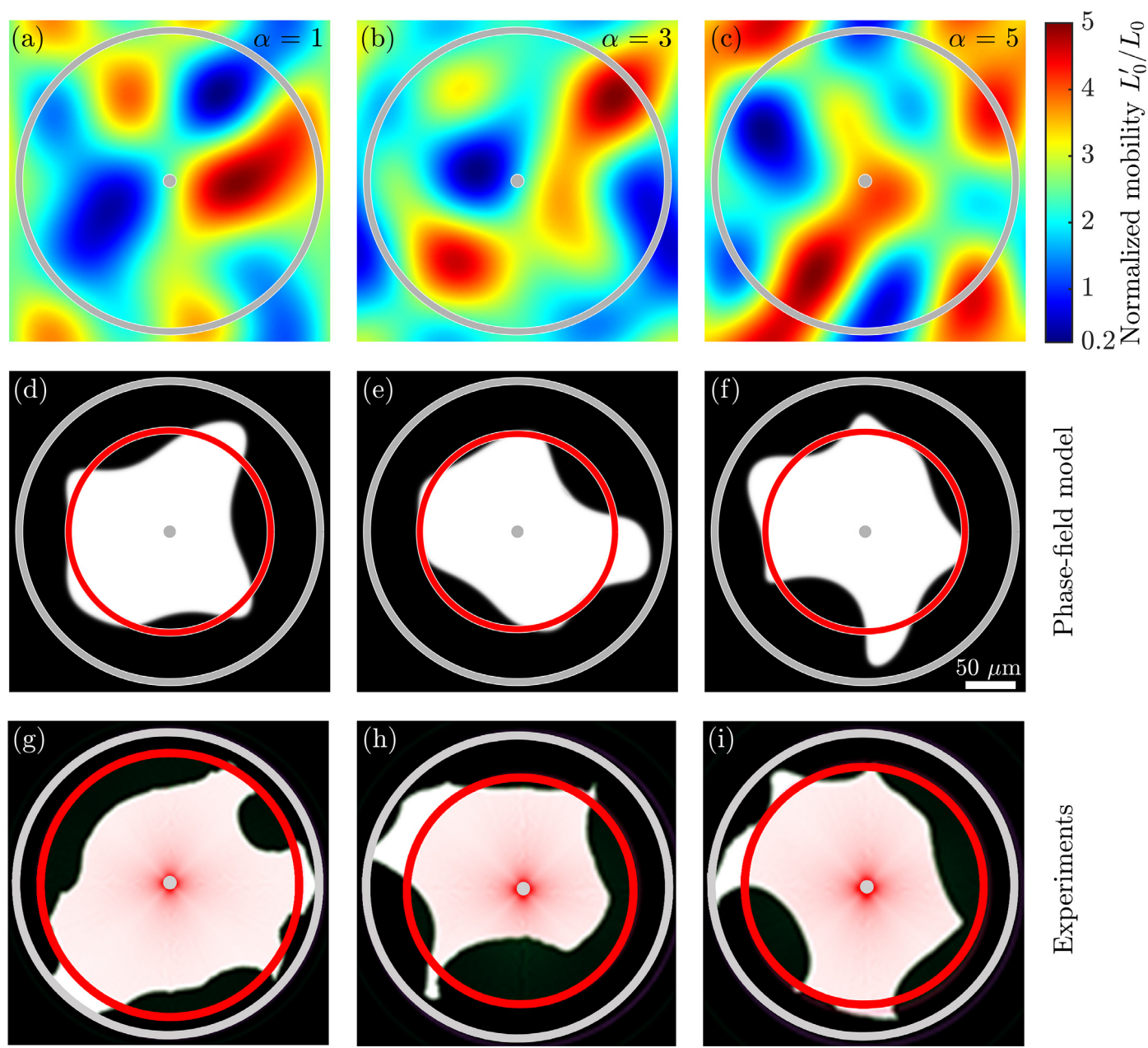

**Fig. 5.** (a–c) Spatial distribution of the mobility parameter $L'_0$ generated with $N=2$ and $\beta=0.1$ for three different $\alpha$ values. (d–f) Phase-field predictions of the cross-section of the Mg wire after 24 h of immersion in SBF using $L'_0$ from (a–c). (g–i) Representative experimental cross-sections of the Mg wires after 24 h of immersion in SBF. The grey points and grey lines indicate the center and initial cross-section of the Mg wire before degradation in both experiments and simulations. The red circle stands for a uniform corrosion radius. The scale bar for all figures is 50 μm.

Mg concentration. The protective film is modeled as an impermeable layer with a thickness of 0.5 μm around the wire surface with the corresponding no flux boundary condition for both Mg concentration and phase-field. To investigate the influence of the mechanical fields on corrosion kinetics, additional constraints are enforced for the mechanical equilibrium equation. The normal component of the displacement vector along the vertical and horizontal symmetry axes is constrained ($\mathbf{n} \cdot \bar{\mathbf{u}} = 0$) while a non-zero remote tensile deformation $\varepsilon^\infty$ is prescribed on the top surface, Fig. 7. The remote deformation is prescribed at the beginning of the simulation and held fixed over a total simulation time. Its magnitude is varied to study the role of mechanical fields in enhancing corrosion kinetics and SCC behavior.

The material properties and the phase-field parameters used are the same as in the previous case study of uniform corrosion. The interfacial mobility coefficient $L$ incorporates the role of mechanical fields through Eq. (20), whereas $L_0$ is previously determined in the comparison with load-free experiments. The mechanical properties of an AZ31 Mg alloy from the literature are used for the simulations [27]. The Mg alloy is assumed to behave as an isotropic, elasto-plastic solid. The Lamé elastic constants are $\lambda = 38$ GPa and $\mu = 16.3$ GPa. Plastic deformation is described using the J$_2$ flow theory with non-linear isotropic hardening, with a yield stress of 138 MPa and an ultimate tensile strength of 245 MPa at an engineering strain of 17%.

The obtained results in terms of phase-field contours, Mg concentration distribution, and mechanical fields for various remote deformations $\varepsilon^\infty$ after 24 h of immersion in SBF are presented in Fig. 8. In the absence of mechanical loading ($\varepsilon^\infty = 0$), the pit grows uniformly, keeping the initial circular shape with a low and uni-form concentration of Mg ions within the pit. The application of a relatively small axial deformation (e.g., $\varepsilon^\infty = 0.096\%$) increases the magnitude of $\sigma_h$ in a small localized area and produces negligible plastic deformation. The hydrostatic stress distribution changes the pit morphology initiating a pit-to-crack transition. The Mg concentration increases near the tip of the pit, indicating that corrosion of Mg is localized in this region because of the stress concentration associated with the sharp tip. Further increase in the applied strain ($\varepsilon^\infty = 0.1\%$) raises the stresses high enough to trigger noticeable plastic deformations. The shape of the evolving defect is governed by both the hydrostatic stress and the plastic strain distribution. Longer and smoother cracks are observed compared to the previous case ($\varepsilon^\infty = 0.096\%$), as the hydrostatic stress and plastic strain distributions engage a more extensive area. The Mg concentration is significantly increased at the crack tip, but it is still well below the equilibrium value in the liquid phase ($c_{Mg}^{l,eq} = 0.57$ mol/L), indicating an activation-controlled process. Model predictions in terms of pit depth and hydrogen gas evolution for the cases considered are given in Fig. 9. Both pit kinetics and hydrogen gas production increase with an increase in applied strain. However, the pit kinetics is dramatically altered in the presence of mechanical loading, leading to rapid crack growth and fracture of the wires after a short time in SBF.

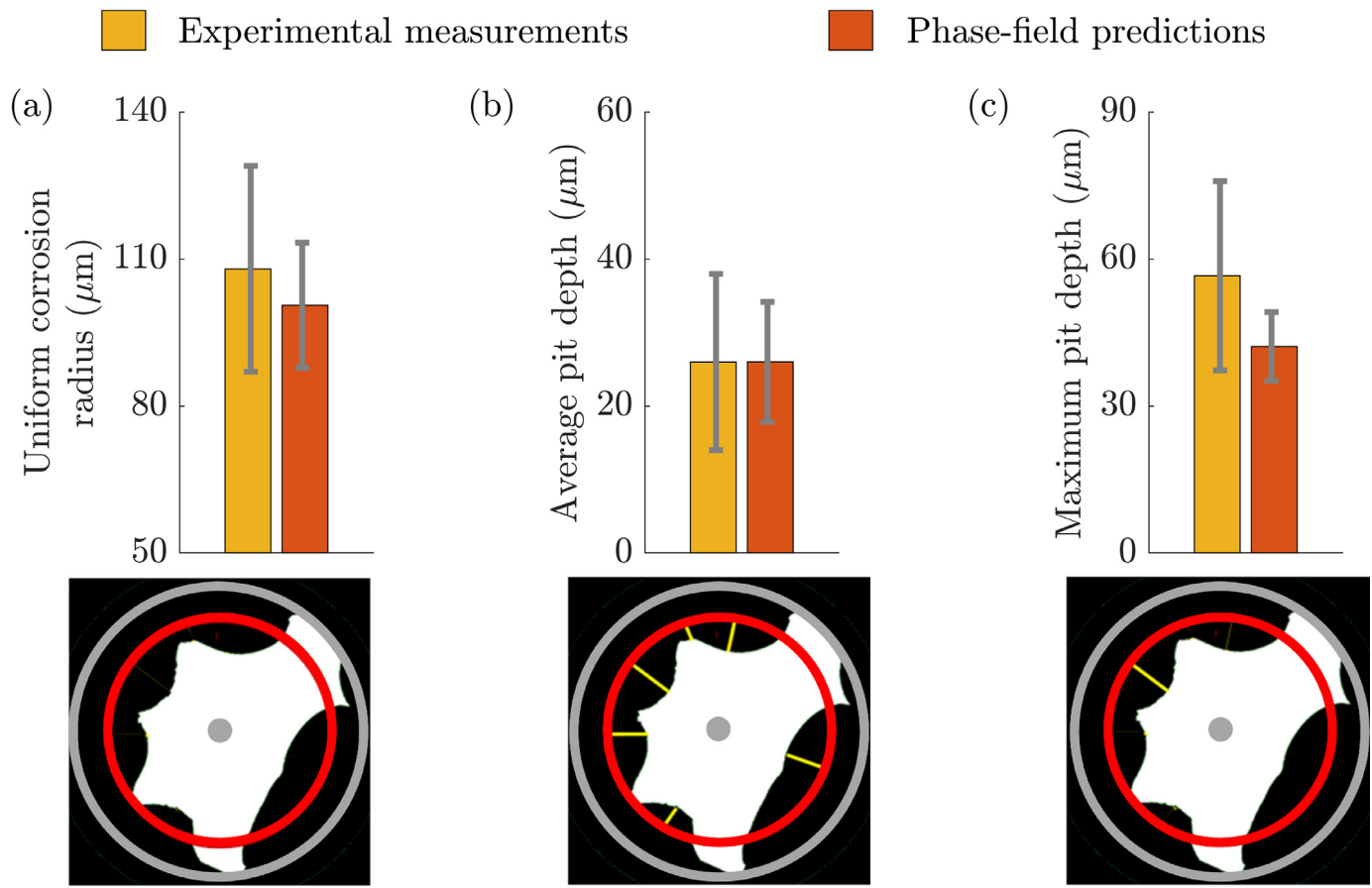

**Fig. 6.** Comparison between experimental and simulated values for the three pitting metrics parameters. (a) Uniform corrosion radius (the radius of the circular section that has the same area as the corroded cross-section). (b) Average pit depth (average distance from the degraded cross-section to the uniform corrosion circle). (c) Maximum pit depth (maximum distance from the corroded cross-section to the uniform corrosion circle).

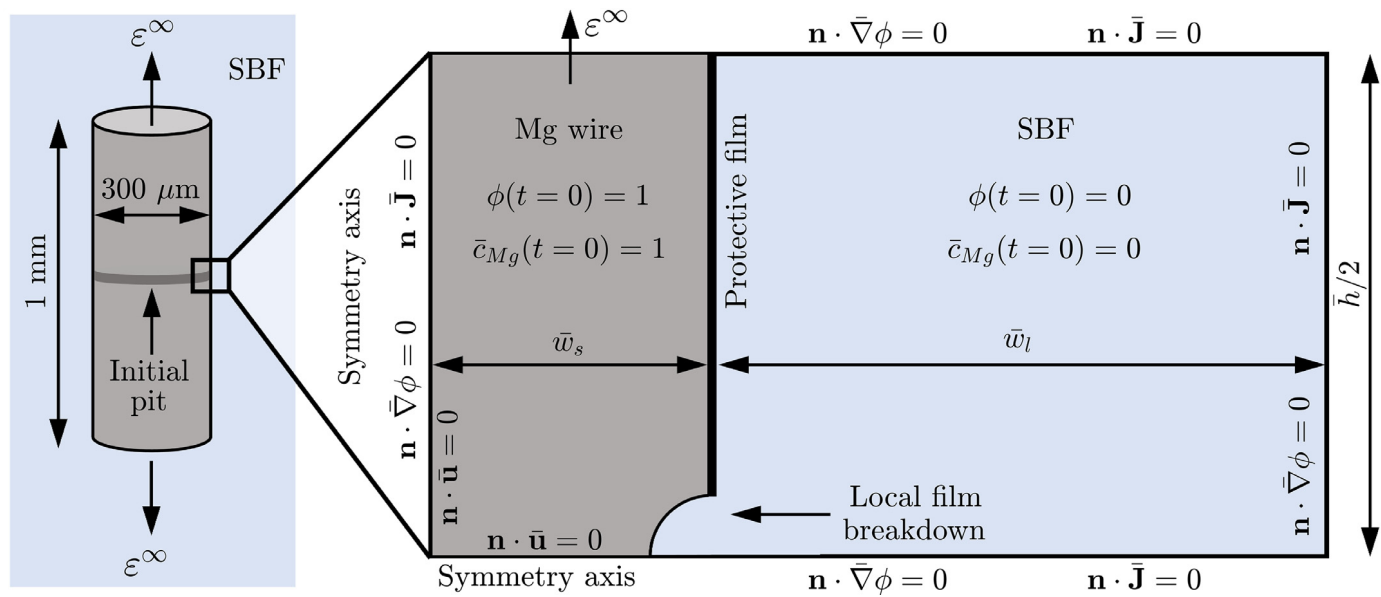

**Fig. 7.** Simulation domain for the Mg alloy wire of 1 mm in length and 300 μm in diameter immersed in SBF and subjected to tensile deformation along the wire axis. The semi-circular pit created by the rupture of the protective layer has a radius of 10 μm. The size of the nondimensional computational domain ($\bar{w}_s = 37.5$, $\bar{w}_l = 362.50$, and $\bar{h} = 250$) is normalized using the interface thickness $\ell = 4$ μm as the characteristic length.

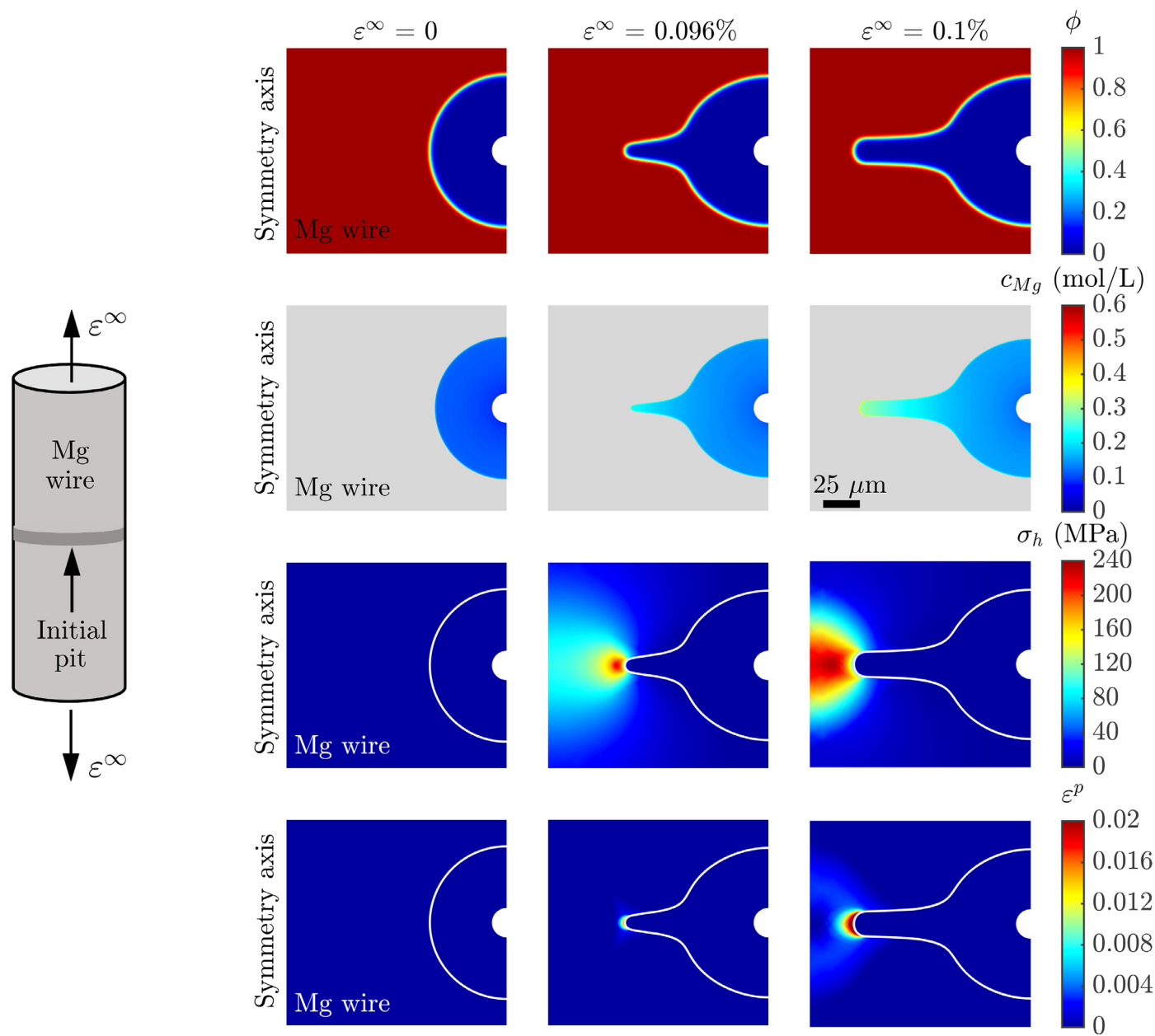

**Fig. 8.** Contour plots of the phase-field variable, Mg concentration distribution in SBF, hydrostatic stress $\sigma_h$, and effective plastic strain $\varepsilon^p$ for various prescribed remote deformations $\varepsilon^\infty$ after 24 h of immersion in SBF. The initial surrounding corrosive environment is not shown in the plots.

### 4.2. Bioabsorbable coronary Mg stent

The potential of the model is demonstrated in predicting the degradation of a bioabsorbable coronary Mg stent immersed in biological fluid. Mg alloy-based stents are attractive as temporary scaffolds to diseased blood vessels and exhibit good clinical performance [5]. However, premature failure due to fast corrosion rates limits their cardiovascular applications. A physically-based model for uniform corrosion [35] and phenomenological approaches [74–76] have been employed in simulating the degradation of Mg stents, considering various stent geometries and Mg alloys. The present phase-field corrosion model is applied in this section to simulate stent degradation taking into account both uniform and pitting corrosion.

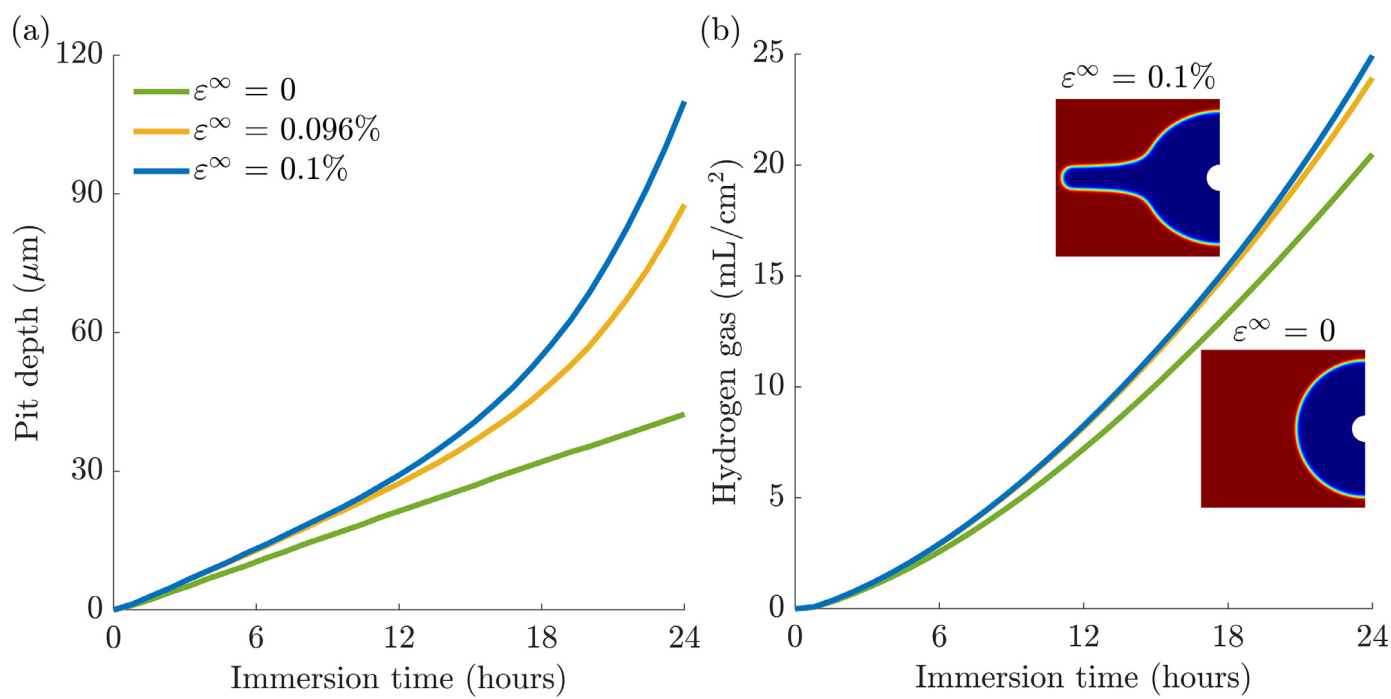

**Fig. 9.** Model predictions of (a) pit depth and (b) hydrogen release as a function of immersion time for different applied axial strains.

An idealized stent geometry that resembles different stent designs frequently used in the literature [74–76] is considered in this work. The stent has an outer diameter of 2 mm and a total length of 7.5 mm. The whole geometry comprises six rings interconnected by a link with a length of 0.30 mm and a diameter of 0.125 mm. Each ring has six peak-to-valley struts with a total height of 1 mm and a diameter of 0.15 mm, Fig. 10(a).

The computational domain consists of the stent and the surrounding corrosive environment. It is assumed that the stent is immersed in a physiologically representative blood vessel environment such that the initial Mg concentration corresponds to human blood (0.875 mmol/L). As in the previous examples, no-flux boundary conditions are imposed on all the external surfaces of the domain for phase-field and Mg concentration. The size of the corrosive environment is significantly larger than the stent geometry to avoid saturation effects. The material properties, the phase-field parameters, and the interfacial mobility parameter $L_0$ follow those used in the previous examples.

Two different case studies are considered. In the first study, the stent is mechanically loaded before immersion in the corrosive environment. It is radially expanded to an outer diameter of 2.25 mm, mimicking the balloon inflation stage during the deployment process. The balloon is modeled as a rigid cylindrical body. This step is followed by the stent recoil, which corresponds to the balloon deflation and extraction process. The final stent outer diameter following the recoil is 2.168 mm. The stent deployment process is summarized in Fig. 10(b). The stress state in terms of von Mises stresses and equivalent plastic strains for the representative ring element after stent recoil is shown in Fig. 10(c). The plastic strains are then incorporated into the subsequent corrosion simulation. In the second study, the as-manufactured stent is immersed in biological fluid in the absence of mechanical stresses. This case corresponds to uniform corrosion and is a reference study for comparison.

The results of the phase-field simulations for mechanically assisted and uniform corrosion are given in Fig. 11. In the former case, plastic strains are localized at the union between rings and links (as shown in Fig. 10(c)), providing hot spots for pitting nucleation. Mass loss ratio (computed using Eq. (28) as $\Delta n_{Mg}/n_{Mg}^{t=0}$) in Fig. 11(a) shows that pitting corrosion is initiated immediately after immersion in SBF due to the initial plastic strains, whereas uniform corrosion progresses more slowly. After 24 h of immersion in SBF, pitting corrosion returns a slightly higher mass loss ratio than uniform corrosion. Although Fig. 11(a) indicates that the stent dissolves faster in the presence of mechanical fields, pitting corrosion notably deteriorates the structural integrity of the stent, as further elaborated. Phase-field isosurface plots after 24 h of immersion in solution for the first case study considered are presented in Fig. 11(b). A pitting zone is observed in the vicinity of the union between rings and links. The dissolution rate within the pitting zone is much higher than in the remaining parts of the stent. This locally enhanced dissolution significantly reduces the thickness of the strut, as shown in two characteristic cross-sections close to the union point in Fig. 11(c), indicating hot spots for early stent failure. The structural integrity of the stent at these locations is severely undermined. In the case of uniform corrosion, the contour plots in Fig. 11(c) show that the stent gradually dissolves, covering the whole sample with a constant dissolution rate. This example demonstrates the importance of including mechanical fields in analyzing the degradation of coronary Mg stents. These structures inevitably experience complex stress states during deployment and service, and thus, uniform corrosion models would give overestimated service life predictions.

## 5. Discussion

The present diffuse interface model for assessing the *in vitro* corrosion of biodegradable Mg-based alloys captures different corrosion mechanisms. Uniform corrosion is included through a constant interfacial mobility parameter calibrated with *in vitro* corrosion data in terms of hydrogen gas evolution (or mass loss). A spatially-dependent interfacial mobility parameter (Eq. (29)) is introduced to simulate pitting corrosion. Its spatial dependence is correlated to the nonuniform distribution of pitting (chloride) ions in the corrosive environment. The model reproduces reasonably well pitting metrics experimentally observed in Mg wires, Figs. 5 and 6. As demonstrated in Fig. 5, the model readily captures complex geometries and geometric interactions such as multiple pits, pit coalescence, and pit growth. It should be emphasized that the present work represents the first physically-based model to simulate pitting corrosion (surface-based localized corrosion) in biodegradable Mg alloys.

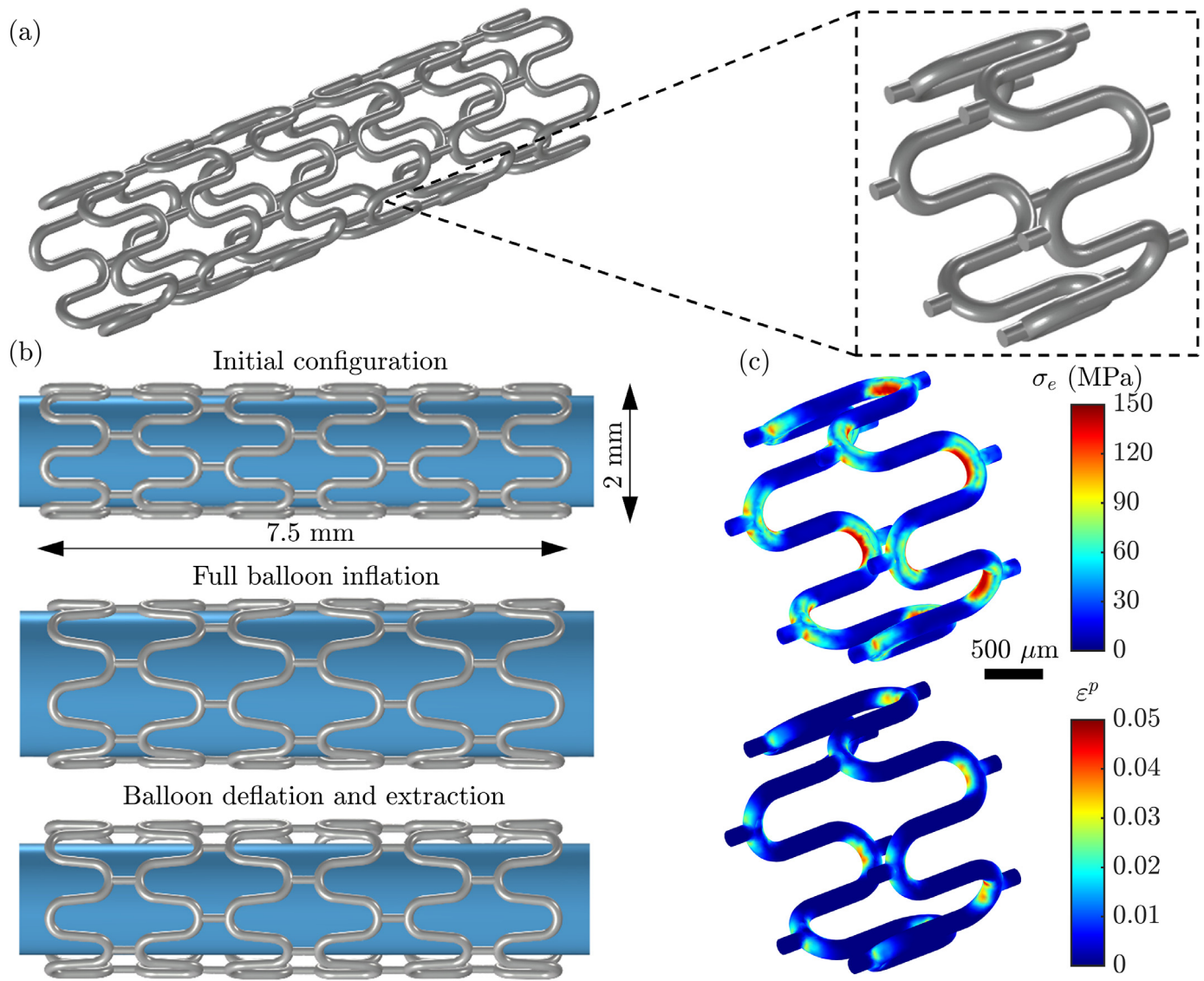

**Fig. 10.** Bioabsorbable coronary Mg stent. (a) Idealized stent geometry and representative ring element. (b) Stent deployment steps. (c) von Mises stresses $\sigma_e$ and equivalent plastic strains $\varepsilon^p$ after stent recoil.

The potential of the model for capturing the role of mechanical fields in enhancing the corrosion of Mg alloys is demonstrated by modeling the behavior of a circumferential sample containing a notch and undergoing tensile testing (Section 4.1). The mechanical contribution is incorporated via a mechano-electrochemical effect that depends on local stress and strain distributions, Section 2.4. Mechanical stresses have deleterious influences on the corrosion resistance of Mg alloys, as previously observed in several *in vitro* studies [19–22], leading to the localization of damage and the formation of sharp cracks that accelerate failure. Changes in pit morphology, increased hydrogen gas production, pit-to-crack transition initiation, and faster crack propagation are noticed under external loads in Figs. 8 and 9. More importantly, the model shows that once the pit-to-crack transition develops, it leads to rapid and uncontrollable crack growth. For practical purposes, the model can be utilized in designing and estimating the service life of load-bearing biomedical devices from Mg alloys. Moreover, it can serve as an effective tool to foresee the mechanical strength of body implants (i.e. scaffolds for bone tissue engineeringas well as fixation devices for bone fracture) after a certain period of degradation depending on their geometry and to preempt catastrophic implant failures.

The proposed framework is also used to assess the effect of complex stress conditions, which arise during stent deployment and service, on the corrosion of bioabsorbable Mg stents. Pitting corrosion, initiated due to local plastic strains developed during stent deployment (Fig. 10(c)), proves to have more detrimental effects on stent degradation than uniform corrosion, Fig. 11. The model may serve as a cost-effective way of predicting the degradation of Mg stents and assessing their residual strength during the degradation process. The availability to foresee the locations of early break points in the sample and determine scaffolding capabilities during degradation is appealing for practical applica-tions in the design of biomedical devices such as bioabsorbable stents. Obtaining an optimized stent design is beyond the scope of the current paper. However, integrating the proposed model with optimization analysis would return more sophisticated stent designs with improved corrosion performance and help develop new bioabsorbable metallic stents.

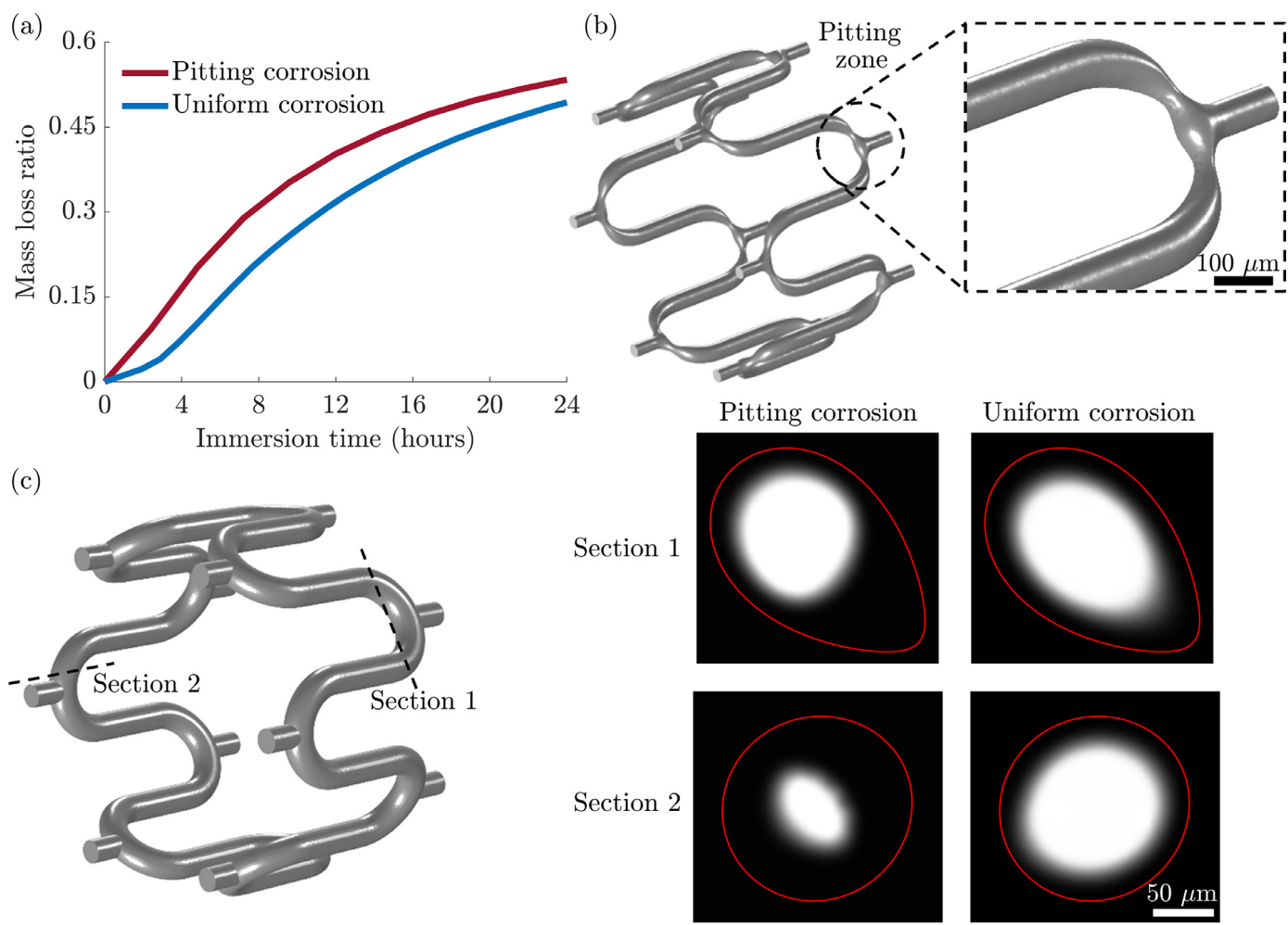

**Fig. 11.** Pitting and uniform corrosion of bioabsorbable coronary Mg stents. (a) Mass loss ratio as a function of immersion time. (b) Phase-field isosurface plots for pitting corrosion after 24 h of immersion. The pitting zone is observed in the areas of high plastic strains. (c) Phase-field contour plots of two characteristic cross-sections after 24 h of immersion. The red line indicates the initial cross-section of the Mg stent before degradation.

The model potential thus being discussed, it is important to address the questions regarding the role of the equilibrium concentration in the liquid phase and the rate-limiting process on the corrosion behavior, showing additional model capabilities. The formation of the partly protective layer is neglected in the present formulation. The saturated concentration of Mg ions in the corrosive environment is determined based on the mass density and molar mass of $MgCl_2$ formed on the exposed Mg surface (Section 2.1). This yields the equilibrium concentration of Mg ions in the liquid phase $c_{Mg}^{l,eq} = 0.57$ mol/L. Taking a higher value for the saturated concentration would lead to difficulties in forming the protective layer, thereby promoting corrosion. On the contrary, decreasing $c_{Mg}^{l,eq}$ would physically represent easier precipitation of the protective layer on the metal surface, increasing corrosion resistance and consequently decelerating the degradation process. To show the effect of $c_{Mg}^{l,eq}$ on the corrosion process, two additional case studies are considered with lower $c' = 0.5 c_{Mg}^{l,eq}$ and higher $c'' = 2 c_{Mg}^{l,eq}$ equilibrium concentrations in the liquid phase while keeping all the other parameters fixed as in Section 4.1. Phase-field contours and Mg concentration distributions for the final pit shape are shown in Fig. 12. As expected, the pit depth increases with the equilibrium concentration in the liquid phase. Hence, the proposed framework can be tweaked to capture the formation of the protective film or other phenomena related to Mg surface modifications by varying $c_{Mg}^{l,eq}$.

The rate-limiting process between diffusion- and activation-controlled corrosion is defined in Eq. (25). Considering the geometry and material properties as in Section 4.1, two corrosion tests are conducted to illustrate the effect of the rate-limiting process on corrosion behavior. Phase-field contours for the final pit shape and Mg concentration distribution in the absence of mechanical load are shown in Fig. 13 for both rate-limiting processes. In agreement with expectations for the diffusion-controlled process ($\tau \gg 1$), the pit growth is pronounced and Mg concentration around the interface is close to the equilibrium value in the liquid phase $c_{Mg}^{l,eq}$. On the contrary, pit growth is slower and Mg concentration stays significantly below $c_{Mg}^{l,eq}$ for the activation-controlled process ($\tau \ll 1$), Fig. 13.

The present phase-field formulation overcomes the limitations in tracking the evolution of corrosion interfaces in arbitrary domains under complex physics and handling complex topological changes without requiring ad hoc criteria. The current paper focuses on biodegradable Mg alloys due to their high attractiveness as biomaterials. However, the framework developed is general and easily extendable to other biodegradable metals, such as Fe and Zn-based alloys, using the corresponding material properties and the composition of the corrosion layers. The advantages of the phase-field method can be further exploited by extending the present formulation and adding other physical phenomena, such as the electrochemistry-corrosion interplay. As emphasized in Section 2.1, the reactions for product formation and layer dissolution (Eqs. (1) and (2)) are neglected in the current model. Thus, the degradation mechanism is based on the anodic reaction and diffusion of Mg ions in the physiological environment. The contribution of the electric field to material dissolution and species diffusion is also neglected in the present model, as the Mg ions are not considered as charged species. These limitations could be overcome by incorporating the reactions for product formation and layer dissolution along with the transport of charged ions, their interactions, and electric field distribution. This would make the model more advantageous and enhance its versatility. Such a model would contribute to understanding the underlying electrochemical process and disclose the effect of the composition of the environment on the corrosion process. The extension to incorporate the electrochemistry-corrosion interplay will be addressed in future works. Incorporating the above-mentioned ingredients could potentially solve the long-term open question of the mismatch of corrosion rates between *in vivo* and *in vitro* tests. In addition, future work should consider the effect of microstructural features, such as alloying elements, grain size/shape, grain boundaries, and interfacial energy dependence on grain orientation, on Mg corrosion. These features would deliver new scientific insight into other phenomena related to Mg corrosion, such as intra- and trans-granular corrosion.

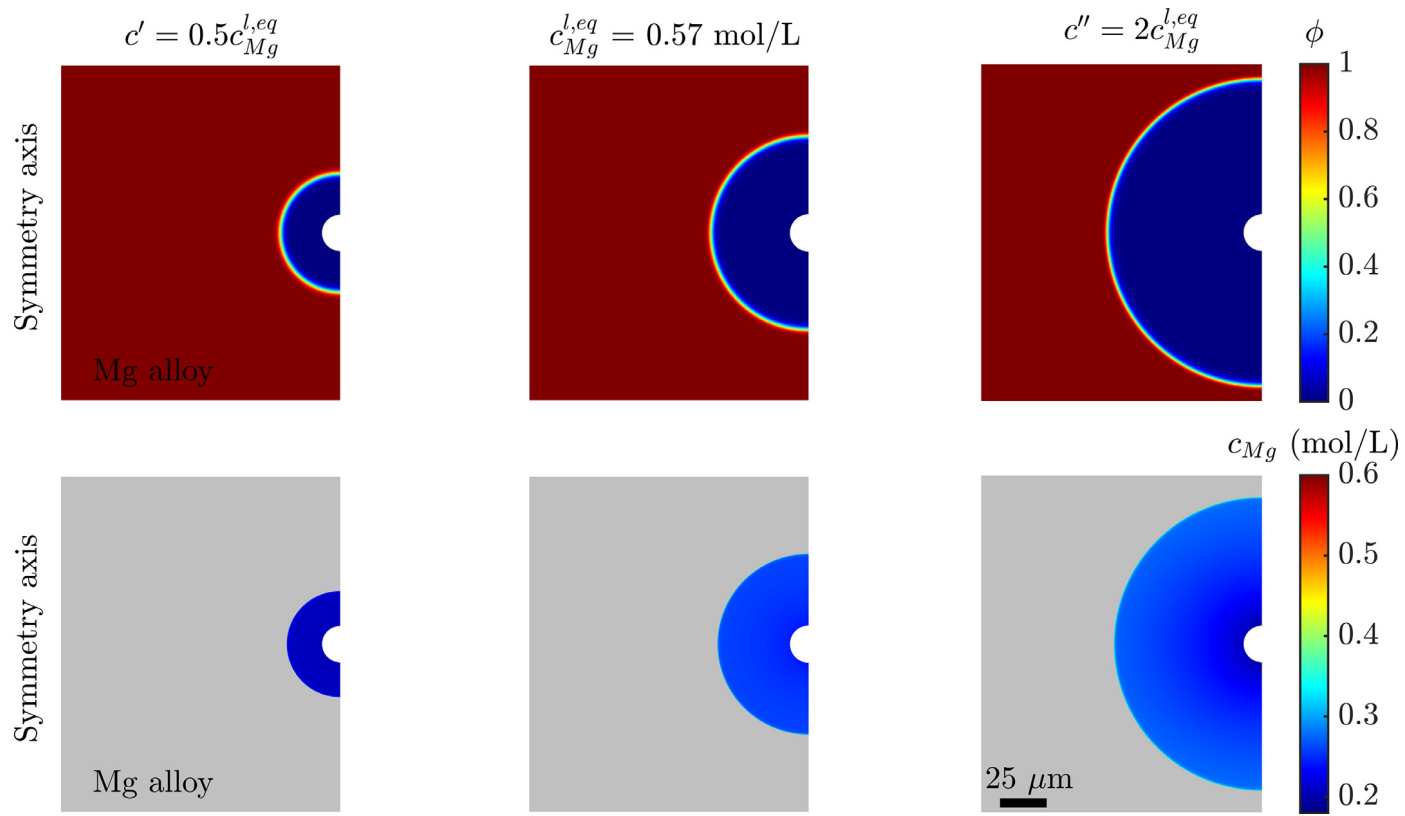

**Fig. 12.** Contour plots of the phase-field variable and Mg concentration distribution in liquid for different equilibrium concentrations of Mg ions in the liquid phase ($c_{Mg}^{l,eq}$) after 24 h of immersion in SBF in the absence of mechanical loading. The simulation domain corresponds to Fig. 7. The surrounding corrosive environment is not shown in the plots.

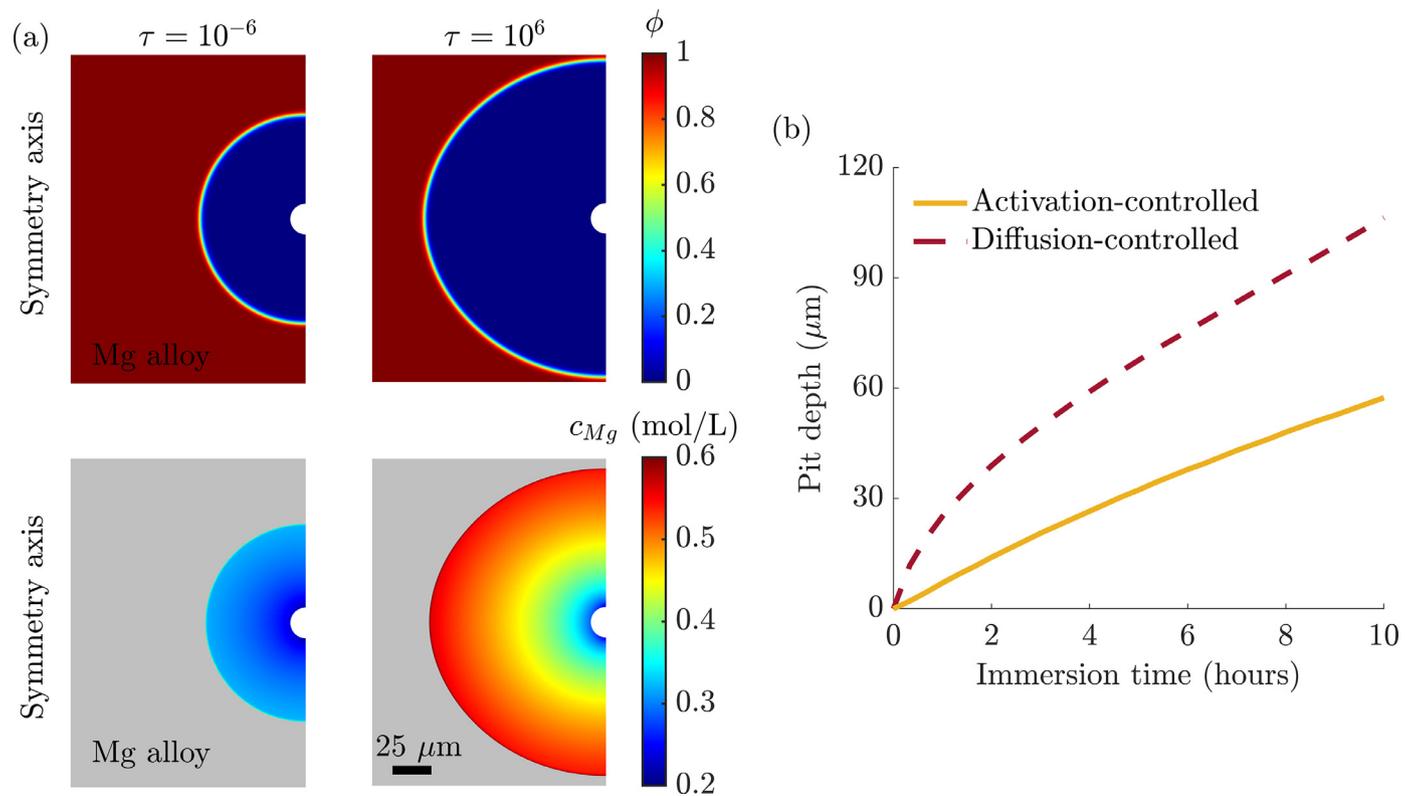

**Fig. 13.** (a) Contour plots of the phase-field variable and Mg concentration distribution in liquid for diffusion-controlled ($\tau = 10^6$) and activation-controlled corrosion ($\tau = 10^{-6}$) after 10 h of immersion in SBF. The simulation domain corresponds to Fig. 7. The surrounding corrosive environment is not shown in the plots. (b) Pit depth as a function of immersion time for the two rate-limiting processes.

## 6. Conclusions

A computational framework based on the phase-field method has been presented for assessing the corrosion of biodegradable Mg alloys in physiological environments that resemble biological media. Built upon thermodynamical principles, the model uses an Allen-Cahn equation to capture Mg dissolution, a diffusion equation to estimate the diffusion of Mg ions in solution, and a mechano-chemical enhancement of the phase-field mobility coefficient to capture the interplay between corrosion kinetics and mechanical fields. In addition to uniform corrosion, pitting corrosion is introduced assuming nonuniform distributions of chloride ions in solution. The proposed framework applies to arbitrary two-dimensional and three-dimensional geometries with no special treatment for the evolution of the corrosion front.

The model parameters for uniform corrosion are calibrated with *in vitro* corrosion data and predictions of pitting corrosion are compared with experiments conducted on Mg wires. A good agreement between experiments and simulations is retrieved. The importance of including the effect of pitting corrosion mechanism and mechanical loads in accelerating degradation is demonstrated in representative case studies: pitting corrosion associated with the local failure of a protective layer and the nonhomogeneous stress state of a bioabsorbable coronary stent. The following conclusions can be drawn:

(i) Mechanical loading significantly alters corrosion kinetics and has deleterious influences on the corrosion resistance of Mg alloys. The application of tensile deformation changes the pit morphology and initiates a pit-to-crack transition. Further increase in mechanical loading may trigger rapid crack growth and premature fracture after a short time in SBF, as previously observed in *in vitro* studies.
(ii) Local plastic strains developed during stent deployment act as initiators for pitting corrosion, indicating hot spots for early stent failure. The results show that pitting corrosion in the stent is initiated immediately after immersion in SBF due to the initial mechanical strains, whereas uniform corrosion progresses more slowly, covering the whole sample with a constant dissolution rate. In addition, pitting corrosion proves to have more detrimental effects on stent degradation than uniform corrosion and severely compromises the structural integrity of the stent. This study reveals that neglecting the mechanical effects on stent degradation and considering uniform corrosion would lead to unsafe design solutions and overestimated service life predictions of coronary Mg stents.

The proposed framework can assist in designing and predicting the service life of biomedical devices after a certain period of immersion. The model may serve as a complementary tool for planning *in vitro,* tests and as a cost-effective way of assessing the residual strength and scaffold capabilities of temporary body implants such as bioabsorbable stents, bone implants or porous scaffolds for tissue regeneration.

## Declaration of Competing Interest

The authors declare that they do not have competing financial interests or personal relationships that could have influenced the work reported in this paper.

## Acknowledgments

W.A. and J.LL. acknowledge financial support from the BIOMET4D project (Smart 4D biodegradable metallic shape-shifting implants for dynamic tissue restoration) under the European Innovation Council Pathfinder Open call, Horizon Europe Research and Innovation Program, grant agreement No. 101047008, and from the Spanish Research Agency through the grant PID2021-124389OB-C21. S.K. and E.M.-P. acknowledge financial support from UKRI's Future Leaders Fellowship program [Grant MR/V024124/1].

## Supplementary material

Supplementary material associated with this article can be found, in the online version, at 10.1016/j.actbio.2023.04.011 The code developed is available at www.imperial.ac.uk/mechanics-materials/codes. The experimental data in Figures 4 and 6 are available at https://doi.org/10.5281/zenodo.7750715.

## References

[1] F. Witte, The history of biodegradable magnesium implants: a review, Acta Biomater. 6 (2010) 1680–1692.
[2] D. Bairagi, S. Mandal, A comprehensive review on biocompatible Mg-based alloys as temporary orthopaedic implants: current status, challenges, and future prospects, JMA 10 (2022) 627–669.
[3] Y. Zheng, X. Gu, F. Witte, Biodegradable metals, Mater. Sci. Eng. R Rep. 77 (2014) 1–34.
[4] H. Windhagen, K. Radtke, A. Weizbauer, J. Diekmann, Y. Noll, U. Kreimeyer, R. Schavan, C. Stukenborg-Colsman, H. Waizy, Biodegradable magnesium-based screw clinically equivalent to titanium screw in hallux valgus surgery: short term results of the first prospective, randomized, controlled clinical pilot study, Biomed. Eng. Online 12 (1) (2013) 62.
[5] H.M. Garcia-Garcia, M. Haude, K. Kuku, A. Hideo-Kajita, A. Ince, A. Abizaid, R. Tölg, P.A. Lemos, C. von Birgelen, E.H. Christiansen, W. Wijns, J. Escaned, J. Dijkstra, R. Waksman, In vivo serial invasive imaging of the second-generation drug-eluting absorbable metal scaffold (Magmaris - DREAMS 2G) in de novo coronary lesions: Insights from the BIOSOLVE-II First-In-Man Trial, Int. J. Cardiol. 255 (2018) 22–28.
[6] D. Xia, F. Yang, Y. Zheng, Y. Liu, Y. Zhou, Research status of biodegradable metals designed for oral and maxillofacial applications: a review, Bioact. Mater. 6 (2021) 4186–4208.
[7] N. Kirkland, N. Birbilis, M. Staiger, Assessing the corrosion of biodegradable magnesium implants: a critical review of current methodologies and their limitations, Acta Biomater. 8 (2012) 925–936.
[8] F. Witte, J. Fischer, J. Nellesen, H.-A. Crostack, V. Kaese, A. Pisch, F. Beckmann, H. Windhagen, In vitro and in vivo corrosion measurements of magnesium alloys, Biomaterials 27 (2006) 1013–1018.

[9] A. Krause, N. von der Höh, D. Bormann, C. Krause, F.-W. Bach, H. Windhagen, A. Meyer-Lindenberg, Degradation behaviour and mechanical properties of magnesium implants in rabbit tibiae, J. Mater. Sci. 45 (2010) 624–632.

[10] S. Fischerauer, T. Kraus, X. Wu, S. Tangl, E. Sorantin, A. Hänzi, J. Löffler, P. Uggowitzer, A. Weinberg, In vivo degradation performance of micro-arc-oxidized magnesium implants: a micro-CT study in rats, Acta Biomater. 9 (2013) 5411–5420.

[11] J.-M. Seitz, K. Collier, E. Wulf, D. Bormann, F.W. Bach, Comparison of the corrosion behavior of coated and uncoated magnesium alloys in an in vitro corrosion environment, Adv. Eng. Mater. 13 (2011) B313–B323.

[12] Y. Xin, T. Hu, P. Chu, In vitro studies of biomedical magnesium alloys in a simulated physiological environment: a review, Acta Biomater. 7 (2011) 1452–1459.

[13] W. Ali, M. Echeverry-Rendón, A. Kopp, C. González, J. LLorca, Strength, corrosion resistance and cellular response of interfaces in bioresorbable poly-lactic acid/Mg fiber composites for orthopedic applications, J. Mech. Behav. Biomed. Mater. 123 (2021) 104781.

[14] C. Rendenbach, H. Fischer, A. Kopp, K. Schmidt-Bleek, H. Kreiker, S. Stumpp, M. Thiele, G. Duda, H. Hanken, B. Beck-Broichsitter, O. Jung, N. Kröger, R. Smeets, M. Heiland, Improved in vivo osseointegration and degradation behavior of PEO surface-modified WE43 magnesium plates and screws after 6 and 12 months, Mater. Sci. Eng. C 129 (2021) 112380.

[15] Y. Luo, C. Zhang, J. Wang, F. Liu, K.W. Chau, L. Qin, J. Wang, Clinical translation and challenges of biodegradable magnesium-based interference screws in ACL reconstruction, Bioact. Mater. 6 (2021) 3231–3243.

[16] H. Hornberger, S. Virtanen, A. Boccaccini, Biomedical coatings on magnesium alloys - a review, Acta Biomater. 8 (2012) 2442–2455.

[17] S. Agarwal, J. Curtin, B. Duffy, S. Jaiswal, Biodegradable magnesium alloys for orthopaedic applications: a review on corrosion, biocompatibility and surface modifications, Mater. Sci. Eng. C 68 (2016) 948–963.

[18] H. Li, J. Wen, Y. Liu, J. He, H. Shi, P. Tian, Progress in research on biodegradable magnesium alloys: a review, Adv. Eng. Mater. 22 (2020) 2000213.

[19] L. Chen, C. Blawert, J. Yang, R. Hou, X. Wang, M.L. Zheludkevich, W. Li, The stress corrosion cracking behaviour of biomedical Mg-1Zn alloy in synthetic or natural biological media, Corros. Sci. 175 (2020) 108876.

[20] L. Choudhary, R. Singh Raman, Magnesium alloys as body implants: Fracture mechanism under dynamic and static loadings in a physiological environment, Acta Biomater. 8 (2012) 916–923.

[21] P. Han, P. Cheng, S. Zhang, C. Zhao, J. Ni, Y. Zhang, W. Zhong, P. Hou, X. Zhang, Y. Zheng, Y. Chai, In vitro and in vivo studies on the degradation of high-purity Mg (99.99wt.%) screw with femoral intracondylar fractured rabbit model, Biomaterials 64 (2015) 57–69.

[22] Y. Gao, L. Wang, L. Li, X. Gu, K. Zhang, J. Xia, Y. Fan, Effect of stress on corrosion of high-purity magnesium in vitro and in vivo, Acta Biomater. 83 (2019) 477–486.

[23] N. Eliaz, Corrosion of metallic biomaterials: a review, Materials 12 (3) (2019) 407.

[24] K. Chen, Y. Lu, H. Tang, Y. Gao, F. Zhao, X. Gu, Y. Fan, Effect of strain on degradation behaviors of WE43, Fe and Zn wires, Acta Biomater. 113 (2020) 627–645.

[25] J. Lemaitre, R. Desmorat, Engineering damage mechanics, Ductile, Creep, Fatigue and Brittle Failure, Springer, 2005.

[26] D. Gastaldi, V. Sassi, L. Petrini, M. Vedani, S. Trasatti, F. Migliavacca, Continuum damage model for bioresorbable magnesium alloy devices - application to coronary stents, J. Mech. Behav. Biomed. Mater. 4 (2011) 352–365.

[27] J. Grogan, B. O'Brien, S. Leen, P. McHugh, A corrosion model for bioabsorbable metallic stents, Acta Biomater. 7 (2011) 3523–3533.

[28] N. Debusschere, P. Segers, P. Dubruel, B. Verhegghe, M.D. Beule, A computational framework to model degradation of biocorrodible metal stents using an implicit finite element solver, Ann. Biomed. Eng. 44 (2016) 382–390.

[29] E. Galvin, D. O'Brien, C. Cummins, B. Mac Donald, C. Lally, A strain-mediated corrosion model for bioabsorbable metallic stents, Acta Biomater. 55 (2017) 505–517.

[30] Y. Gao, L. Wang, X. Gu, Z. Chu, M. Guo, Y. Fan, A quantitative study on magnesium alloy stent biodegradation, J. Biomech. 74 (2018) 98–105.

[31] S. Ma, B. Zhou, B. Markert, Numerical simulation of the tissue differentiation and corrosion process of biodegradable magnesium implants during bone fracture healing, ZAMM - J. Appl. Math. Mech. 98 (2018) 2223–2238.

[32] K. van Gaalen, C. Quinn, F. Benn, P.E. McHugh, A. Kopp, T.J. Vaughan, Linking the effect of localised pitting corrosion with mechanical integrity of a rare earth magnesium alloy for implant use, Bioact. Mater. 21 (2023) 32–43.

[33] J. Sanz-Herrera, E. Reina-Romo, A. Boccaccini, In silico design of magnesium implants: macroscopic modeling, J. Mech. Behav. Biomed. Mater. 79 (2018) 181–188.

[34] M. Marvi-Mashhadi, W. Ali, M. Li, C. González, J. LLorca, Simulation of corrosion and mechanical degradation of additively manufactured Mg scaffolds in simulated body fluid, J. Mech. Behav. Biomed. Mater. 126 (2022) 104881.

[35] J. Grogan, S. Leen, P. McHugh, A physical corrosion model for bioabsorbable metal stents, Acta Biomater. 10 (2014) 2313–2322.

[36] Z. Shen, M. Zhao, D. Bian, D. Shen, X. Zhou, J. Liu, Y. Liu, H. Guo, Y. Zheng, Predicting the degradation behavior of magnesium alloys with a diffusion-based theoretical model and in vitro corrosion testing, J. Mater. Sci. Technol. 35 (2019) 1393–1402.

[37] B. Zeller-Plumhoff, T. AlBaraghtheh, D. Höche, R. Willumeit-Römer, Computational modelling of magnesium degradation in simulated body fluid under physiological conditions, J. Magnes. Alloy. (2021).

[38] P. Bajger, J. Ashbourn, V. Manhas, Y. Guyot, K. Lietaert, L. Geris, Mathematical modelling of the degradation behaviour of biodegradable metals, Biomech. Model. Mechanobiol. 16 (2017) 227–238.

[39] A.-K. Gartzke, S. Julmi, C. Klose, A.-C. Waselau, A. Meyer-Lindenberg, H.J. Maier, S. Besdo, P. Wriggers, A simulation model for the degradation of magnesium-based bone implants, J. Mech. Behav. Biomed. Mater. 101 (2020) 103411.

[40] M. Barzegari, D. Mei, S.V. Lamaka, L. Geris, Computational modeling of degradation process of biodegradable magnesium biomaterials, Corros. Sci. 190 (2021) 109674.

[41] A. Hermann, A. Shojaei, D. Steglich, D. Höche, B. Zeller-Plumhoff, C.J. Cyron, Combining peridynamic and finite element simulations to capture the corrosion of degradable bone implants and to predict their residual strength, Int. J. Mech. Sci. 220 (2022) 107143.

[42] L.Q. Chen, Phase-field models for microstructure evolution, Annu. Rev. Mater. Res. 32 (2002) 113–140.

[43] W. Mai, S. Soghrati, R.G. Buchheit, A phase field model for simulating the pitting corrosion, Corros. Sci. 110 (2016) 157–166.

[44] W. Mai, S. Soghrati, A phase field model for simulating the stress corrosion cracking initiated from pits, Corros. Sci. 125 (2017) 87–98.

[45] C. Cui, R. Ma, E. Martínez-Pañeda, A phase field formulation for dissolution–driven stress corrosion cracking, J. Mech. Phys. Solids 147 (2021) 104254.

[46] C. Lin, H. Ruan, S.Q. Shi, Phase field study of mechanico-electrochemical corrosion, Electrochim. Acta 310 (2019) 240–255.

[47] C. Lin, H. Ruan, Multi-phase-field modeling of localized corrosion involving galvanic pitting and mechano-electrochemical coupling, Corros. Sci. 177 (2020) 108900.

[48] C. Lin, H. Ruan, Phase-field modeling of mechano-chemical-coupled stress-corrosion cracking, Electrochim. Acta 395 (2021) 139196.

[49] T. Ansari, Z. Xiao, S. Hu, Y. Li, J. Luo, S. Shi, Phase-field model of pitting corrosion kinetics in metallic materials, Npj Comput. Mater. 38 (2018) 382–390.

[50] T.-T. Nguyen, J. Bolivar, Y. Shi, J. Réthoré, A. King, M. Fregonese, J. Adrien, J.-Y. Buffière, M.C. Baietto, A phase field method for modeling anodic dissolution induced stress corrosion crack propagation, Corros. Sci 132 (2018) 146–160.

[51] C. Xie, S. Bai, X. Liu, M. Zhang, J. Du, Stress-corrosion coupled damage localization induced by secondary phases in bio-degradable Mg alloys: phase-field modeling, J. Magnes. Alloy. (2022).

[52] Y. Xin, K. Huo, H. Tao, G. Tang, P.K. Chu, Influence of aggressive ions on the degradation behavior of biomedical magnesium alloy in physiological environment, Acta Biomater. 4 (2008) 2008–2015.

[53] D. Mei, S.V. Lamaka, X. Lu, M.L. Zheludkevich, Selecting medium for corrosion testing of bioabsorbable magnesium and other metals - a critical review, Corros. Sci. 171 (2020) 108722.

[54] L. Li, B. Liu, R. Zeng, S. Li, F. Zhang, Y. Zou, H.J.X. Chen, S. Guan, Q. Liu, In vitro corrosion of magnesium alloy AZ31 - a synergetic influence of glucose and Tris, Front. Mater. Sci. 12 (2018) 184–197.

[55] Y. Xin, T. Hu, P.K. Chu, Degradation behaviour of pure magnesium in simulated body fluids with different concentrations of $HCO_3^-$, Corros. Sci. 53 (2011) 1522–1528.

[56] C. Ning, L. Zhou, Y.Z.Y. Li, P. Yu, S. Wang, T. He, W. Li, G. Tan, Y. Wang, C. Mao, Influence of surrounding cations on the surface degradation of magnesium alloy implants under a compressive pressure, Langmuir 31 (2015) 13561–13570.

[57] J. Gonzalez, S. Lamaka, D. Mei, N. Scharnagl, F. Feyerabend, M. Zheludkevich, R. Willumeit-Römer, Mg biodegradation mechanism deduced from the local surface environment under simulated physiological conditions, Adv. Healthc. Mater. 10 (2021) e2100053.

[58] G. Liu, J. Han, Y. Li, Y. Guo, X. Yu, S. Yuan, Z. Nie, C. Tan, C. Guo, Effects of inorganic ions, organic particles, blood cells, and cyclic loading on in vitro corrosion of MgAl alloys, J. Magnes. Alloy (2021), doi:10.1016/j.jma.2021.08.034.

[59] Y. Zheng, Y. Li, J. Chen, Z. Zou, Effects of tensile and compressive deformation on corrosion behaviour of a Mg-Zn alloy, Corros. Sci. 90 (2015) 445–450.

[60] E. Gutman, Mechanochemistry of Materials, Cambridge International Science Publishing, Cambridge, UK, 1988.

[61] S.G. Kim, W.T. Kim, T. Suzuki, Phase-field model for binary alloys, Phys. Rev. E 60 (1999) 7186–7197.

[62] S. Kovacevic, R. Pan, D.P. Sekulic, S.Dj. Mesarovic, Interfacial energy as the driving force for diffusion bonding of ceramics, Acta Mater. 186 (2020) 405–414.

[63] J.C. Simo, T.J.R. Hughes, Computational Inelasticity, Springer, 1998.

[64] S.Dj. Mesarovic, Physical foundations of mesoscale continua, in: S.Dj. Mesarovic, S. Forest, H.M. Zbib (Eds.), Mesoscale Models: From Micro-Physics to Macro-Interpretation, CISM International Center For Mechanical Sciences Book Series, Springer, 2019.

[65] S.M. Allen, J.W. Cahn, A microscopic theory for antiphase boundary motion and its application to antiphase domain coarsening, Acta Metall. 27 (1979) 1085–1095.

[66] W. Ali, M. Li, L. Tillmann, T. Mayer, C. González, J. LLorca, A. Kopp, Bioabsorbable WE43 Mg alloy wires modified by continuous plasma-electrolytic oxidation for implant applications. Part I: processing, microstructure and mechanical properties, Biomater. Adv. 146 (2023) 213314.

[67] W. Ali, M. Echeverry-Rendón, G. Domínguez, K.v Gaalen, A. Kopp, C. González, J. LLorca, Bioabsorbable WE43 Mg alloy wires modified by continuous plasma electrolytic oxidation for implant applications. Part II: degradation and biological performance, Biomater. Adv. 147 (2023) 213325.

[68] K. van Gaalen, F. Gremse, F. Benn, P.E. McHugh, A. Kopp, T.J. Vaughan, Automated ex-situ detection of pitting corrosion and its effect on the mechanical integrity of rare earth magnesium alloy - WE43, Bioact. Mater. 8 (2022) 545–558.

[69] B. Hallstedt, Molar volumes of Al, Li, Mg and Si, Calphad 31 (2007) 292–302.
[70] B.-Q. Fu, W. Liu, Z.L. Li, Calculation of the surface energy of hcp-metals with the empirical electron theory, Appl. Surf. Sci. 255 (2009) 9348–9357.
[71] H.L. Skriver, N.M. Rosengaard, Surface energy and work function of elemental metals, Phys. Rev. B 46 (1992) 7157–7168.
[72] J. Gonzalez, R.Q. Hou, E.P. Nidadavolu, R. Willumeit-Römer, F. Feyerabend, Magnesium degradation under physiological conditions - best practice, Bioact. Mater. 3 (2018) 174–185.
[73] B. Zeller-Plumhoff, M. Gile, M. Priebe, H. Slominska, B. Boll, B. Wiese, T. Würger, R. Willumeit-Römer, R.H. Meißner, Exploring key ionic interactions for magnesium degradation in simulated body fluid - a data-driven approach, Corros. Sci. 182 (2021) 109272.
[74] W. Wu, L. Petrini, D. Gastaldi, T. Villa, M. Vedani, E. Lesma, B. Previtali, F. Migliavacca, Finite element shape optimization for biodegradable magnesium alloy stents, Ann. Biomed. Eng. 38 (2010) 2829–2840.
[75] W. Wu, D. Gastaldi, K. Yang, L. Tan, L. Petrini, F. Migliavacca, Finite element analyses for design evaluation of biodegradable magnesium alloy stents in arterial vessels, Mater. Sci. Eng. B 176 (2011) 1733–1740.
[76] J.A. Grogan, S.B. Leen, P.E. McHugh, Optimizing the design of a bioabsorbable metal stent using computer simulation methods, Biomaterials 34 (2013) 8049–8060.